%% file: H1821.tex
\documentclass[usenatbib]{mn2e}

\pdfoutput=1
\usepackage{mathptmx}
\usepackage{amsmath}
\usepackage{url}
\usepackage{times}
\usepackage{array}
\usepackage[pdftex]{graphicx}

\input{defn}

\title[The luminous cluster underlying H1821+643] {The X-ray luminous
  cluster underlying the bright radio-quiet quasar H1821+643}
\author[H.R. Russell et al.]
{\parbox[]{7.in}{H.~R. Russell$^1$\thanks{E-mail: hrr27@ast.cam.ac.uk},
    A.~C. Fabian$^1$, J.~S. Sanders$^1$, R.~M. Johnstone$^1$, K.~M. Blundell$^2$,
    \\W.~N. Brandt$^3$ and C.~S. Crawford$^1$\\
    \footnotesize
    $^1$ Institute of Astronomy, Madingley Road, Cambridge CB3 0HA\\
    $^2$ University of Oxford, Department of Physics, Keble Road,
    Oxford OX1 3RH \\
    $^3$ Department of Astronomy \& Astrophysics, The Pennsylvania
    State University, University Park, Pennsylvania 16802, USA 
  }
}

\begin{document}

\maketitle

\begin{abstract}
  We present a \emph{Chandra} observation of the only low redshift,
  $z=0.299$, galaxy cluster to contain a highly luminous radio-quiet
  quasar, H1821+643.  By simulating the quasar PSF, we subtract the
  quasar contribution from the cluster core and determine the physical
  properties of the cluster gas down to $3\arcsec$ ($15\kpc$) from the
  point source.  The temperature of the cluster gas decreases from
  $9.0\pm0.5\keV$ down to $1.3\pm0.2\keV$ in the centre, with a short
  central radiative cooling time of $1.0\pm0.1\Gyr$, typical of a
  strong cool-core cluster.  The X-ray morphology in the central
  $100\kpc$ shows extended spurs of emission from the core, a small
  radio cavity and a weak shock or cold front forming a semi-circular
  edge at $\sim15\arcsec$ radius.  The quasar bolometric luminosity
  was estimated to be $\sim2\times10^{47}\ergps$, requiring a mass
  accretion rate of $\sim40\Msunpyr$, which corresponds to half the
  Eddington accretion rate.  We explore possible accretion mechanisms
  for this object and determine that Bondi accretion, when boosted by
  Compton cooling of the accretion material, could provide a
  significant source of the fuel for this outburst.  We consider
  H1821+643 in the context of a unified AGN accretion model and, by
  comparing H1821+643 with a sample of galaxy clusters, we show that
  the quasar has not significantly affected the large-scale cluster
  gas properties.
\end{abstract}


\section{Introduction}
The central radiative cooling time of the Intracluster Medium (ICM) in
nearby galaxy clusters can drop below $1\Gyr$ and without a
compensating source of heat the gas would rapidly cool down to low
temperatures (for a review see \citealt{Fabian94}).  However, the
large reservoir of cool gas implied by this cooling flow model
exceeds the amount of molecular gas observed
(eg. \citealt{EdgeFrayer03}), and the inferred star formation rates by
an order of magnitude (\citealt{Johnstone87};
\citealt{HicksMushotzky05}; \citealt{Rafferty06}; \citealt{ODea08}).
In addition, recent high resolution X-ray spectroscopy has been unable to
find the emission signatures of gas cooling below about $1\keV$ at the
extreme rates predicted by the cooling flow model in the absence of
heating (\citealt{Peterson03}; \citealt{Kaastra04};
\citealt{PetersonFabian06}).  A heating mechanism is therefore
required to stabilise the cooling in cluster cores (for a review see
\citealt{McNamaraNulsen07}).

High resolution images of nearby cluster cores taken by the
\emph{Chandra X-ray Observatory} have revealed complex structures,
such as cavities, shock fronts and ripples, which have been produced
by the central AGN (eg. \citealt{Boehringer93}; \citealt{FabianPer00},
\citeyear{FabianPer03}, \citeyear{FabianPer06}; \citealt{McNamara00},
\citeyear{McNamara01}; \citealt{Schindler01}; \citealt{FormanM8705};
\citealt{McNamaraNulsen07}).  X-ray cavities, or bubbles, expanded
into the ICM by the AGN provide a direct and relatively reliable means
of measuring the mechanical energy injected by the SMBH
(eg. \citealt{JonesDeYoung05}).  In this way, AGN heating has been
shown to be energetically capable of balancing the cooling losses in
cluster cores (\citealt{Birzan04}; \citealt{Rafferty06};
\citealt{DunnFabian06}; \citealt{McNamara06}).  Repeated cycles of AGN
mechanical heating could support the cluster gas in a relatively
stable state.

The galaxy cluster surrounding the quasar H1821+643, at redshift
$z=0.299$, was discovered optically by \citet{Schneider92} and later
detected in X-rays by \citet{Hall97} and \citet{Saxton97}.  The
cluster is optically rich with an Abell richness class $>2$
(\citealt{Lacy92}).  The quasar's radio luminosity at $5\GHz$ lies in
the classification for `radio-quiet' quasars (at only
$10^{23.9}\WpHzpsr$) and its $151\MHz$ radio luminosity, which is
$10^{25.3}\WpHzpsr$ (\citealt{Blundell01}), is at the boundary
observed to separate FR I and FR II structures.  A deep VLA radio
image of H1821+643 revealed that, although classified as a radio-quiet
quasar, H1821+643 hosts a giant $300\kpc$ FR I radio source
(\citealt{Blundell01}).  H1821+643 has previously been observed
several times with the \emph{Chandra X-ray Observatory} but in all
cases with the gratings in place (HETG \citealt{Fang02}, LETG
\citealt{Mathur03}) which have prevented a detailed analysis of the
cluster emission.  \citet{Fang02} were able obtain a spectrum
of the cluster from the zeroth-order observation and found a
temperature of $\sim10\keV$ and a metal abundance of $0.3\Zsun$.

In this work, we
present new \emph{Chandra} observations of H1821+643 without the
gratings which have allowed us to analyse the underlying ICM and the
interactions with the powerful central quasar.  By comparing our
results with a sample of nearby cool core clusters, we have explored
the implications of two different modes of AGN feedback for H1821+643
and investigated the impact this would have on the evolution of the
ICM.

We assume $H_0 = 70\kmpspMpc$, $\Omega_m=0.3$ and
$\Omega_\Lambda=0.7$, translating to a scale of $4.4\kpc$ per arcsec
at the redshift $z=0.299$ of H1821+643.  All errors are $1\sigma$
unless otherwise noted.

\section{Data Preparation}
H1821+643 was observed by \emph{Chandra} for a total of $90\ks$ split
into four separate observations taken over nine days with the ACIS-S
instrument (Observation IDs: 9398 $34\ks$, 9845 $25\ks$, 9846 $18\ks$
and 9848 $11\ks$).  The data were analysed using CIAO version 4.1.2
with CALDB version 4.1.2 provided by the \emph{Chandra} X-ray Center
(CXC).  The level 1 event files were reprocessed to apply the latest
gain and charge transfer inefficiency correction and then filtered for
bad grades to remove cosmic ray events.  For this analysis, the
improved background screening provided by VFAINT mode was not applied.
In the pileup affected region some of the real X-ray events would be
marked as background events and excluded by the VFAINT filter leaving
a hole in the quasar surface brightness.  To filter the datasets for
periods affected by flares we examined the ACIS-S1 chip, which did not
contain any extended cluster emission.  The background lightcurve for
this chip was filtered using the \textsc{lc\_clean}
script\footnote{See http://cxc.harvard.edu/contrib/maxim/acisbg/}
provided by M. Markevitch and the $\sim4\ks$ affected by flares were
then excluded from the ACIS-S3 level 2 event files.  The final cleaned
exposure time was $85\ks$.

\begin{figure}
\centering
\includegraphics[width=0.98\columnwidth]{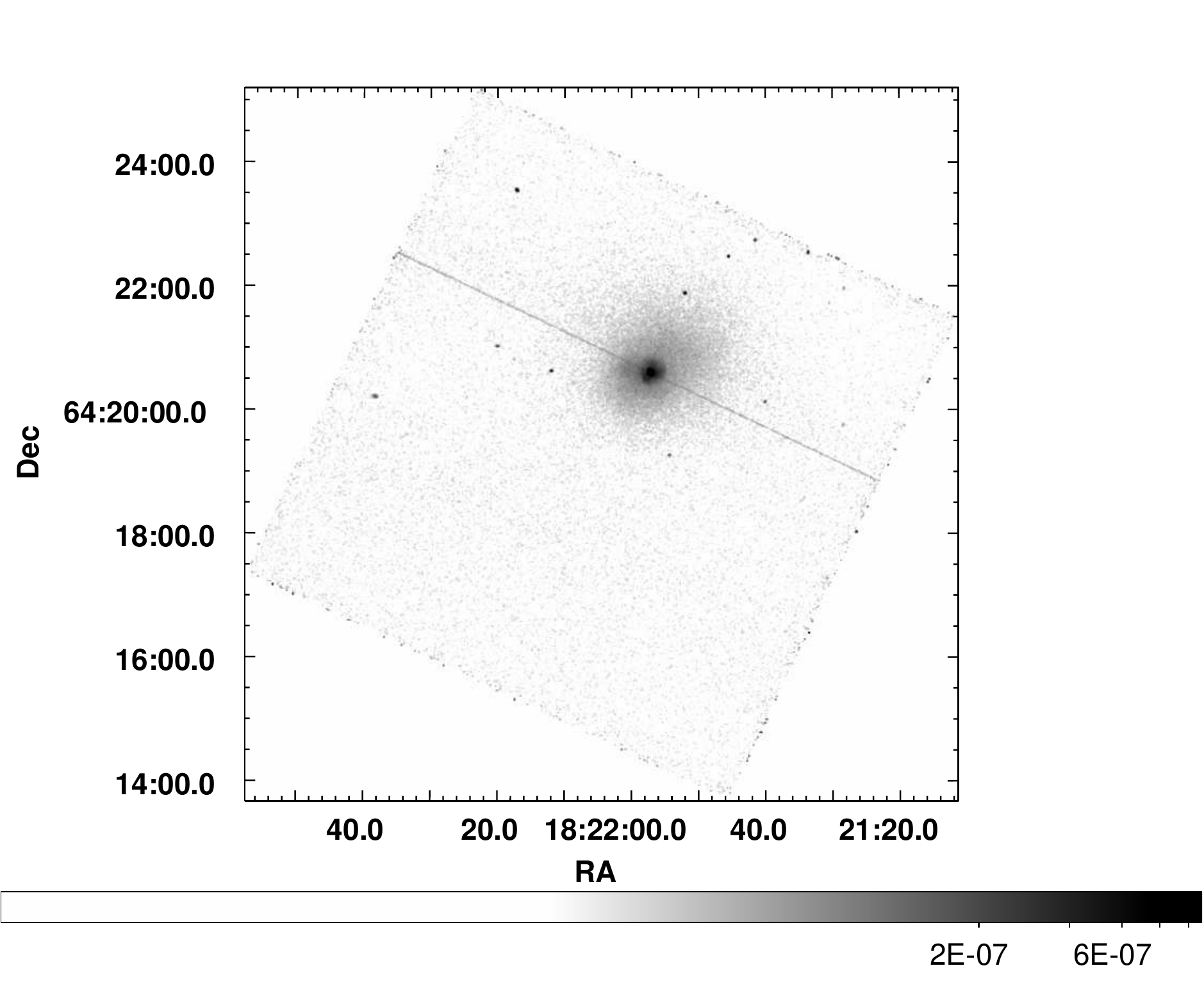}
\caption{Exposure-corrected image of the ACIS-S3 chip in the
  0.5--7.0$\keV$ energy band smoothed with a 2D Gaussian
  $\sigma=1.5\arcsec$.  The logarithmic scale bar has units $\expmap$.
  The straight line running across the entire image through the centre
  of the quasar is the readout streak, an instrumental artifact
  produced by the bright quasar.}
\label{fig:H1821obs}
\end{figure}

As the four separate observations were taken so closely together, with
effectively identical chip positions and roll angles, we were able to
reproject them to a common position (Obs ID: 9398) and combine them.
An exposure-corrected image combining the four final cleaned event
files is shown in Fig. \ref{fig:H1821obs}.  The exposure map assumed a
monoenergetic distribution of source photons of $1.5\keV$, which is
approximately the peak energy of the source.  Assuming that all X-ray
counts within 3\arcsec radius of the point source were from the quasar
and all counts external to that were from the cluster, we estimate
there to be $\sim100,000$ cluster counts and $\sim15,000$ quasar
counts, although the latter will have been significantly reduced by
pileup.

\subsection{Pileup}
Pileup occurs whenever two or more photons, arriving in the same
detector region and within a single ACIS frame integration time, are
detected as a single event (see eg. \citealt{Davis01}).  This effect
hardens the source spectrum, because photon energies sum to create a
detected event of higher energy, and causes grade migration.  All
events detected by ACIS are graded based on the shape of their charge
cloud distributions in a $3 \times 3$ pixel island.  This grade is
used to distinguish between a real photon or a background event, such
as a cosmic ray.  However, pileup modifies the distribution of charge
and therefore alters the event grade.

\begin{figure}
\centering
\includegraphics[width=0.9\columnwidth]{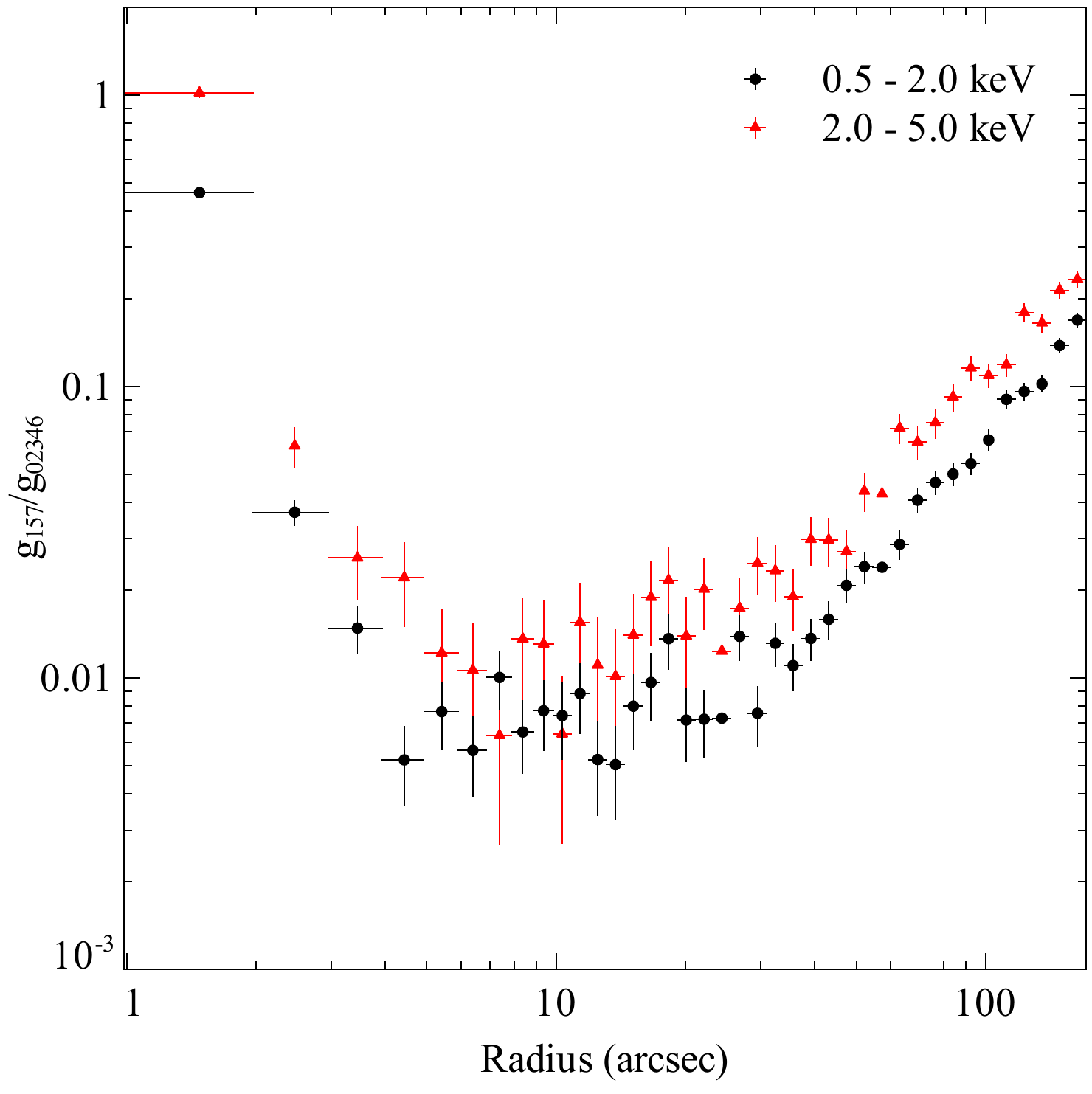}
\caption{Ratio of \emph{ASCA} bad grades (g$_{157}$) to \emph{ASCA}
  good grades (g$_{02346}$) for H1821+643 in two energy bands.}
\label{fig:pileup}
\end{figure}

Following the CXC analysis of the \emph{Chandra} PSF\footnote{See
  http://cxc.harvard.edu/cal/Hrma/psf/}, the extent of pileup in the
observation was estimated using the ratio of ``good''
grades (grades 0,2,3,4,6) to ``bad'' grades (grades 1,5,7) defined
with observations from the \emph{ASCA} satellite.  The
bad/good ratio is shown as a function of radius in Fig.
\ref{fig:pileup} for two energy bands.  The ratio was determined in
circular annuli which scale logarithmically in width to produce
approximately the same number of source counts in each radial bin.  

Most of the emission from the
quasar comes from the $0.5-2.0\keV$ energy band which, when piled up,
was detected as higher energy photons above $2\keV$.  The
$2.0-5.0\keV$ energy band was therefore used to estimate the radial
extent of pileup for this observation.  Fig. \ref{fig:pileup} shows
that pileup rapidly increases inside $3''$ and dominates at $1''$.
The rise in the bad/good ratio beyond $\sim50''$ is caused by the
increasing importance of the particle background.  To minimise the
effects of pileup, the region inside $3''$ was excluded from our
analysis.  This choice was also later confirmed to be appropriate by
repeating the \textsc{marx} simulation of the quasar with pileup
(Appendix A2).

\section{Imaging Analysis}
\label{sec:imaging}
\begin{figure*}
\begin{minipage}[t]{\textwidth}
\centering
\includegraphics[width=0.48\columnwidth]{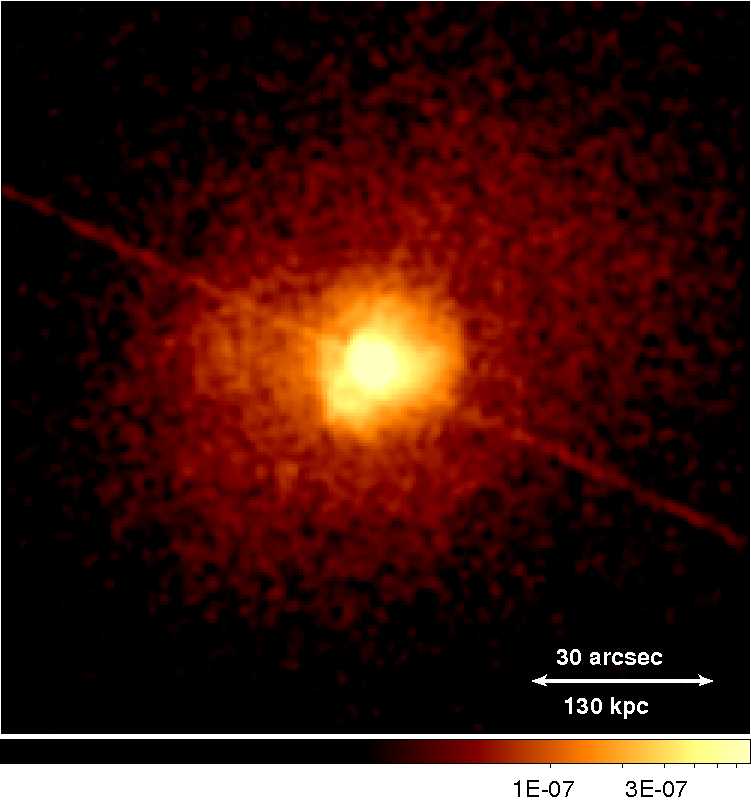}
\includegraphics[width=0.48\columnwidth]{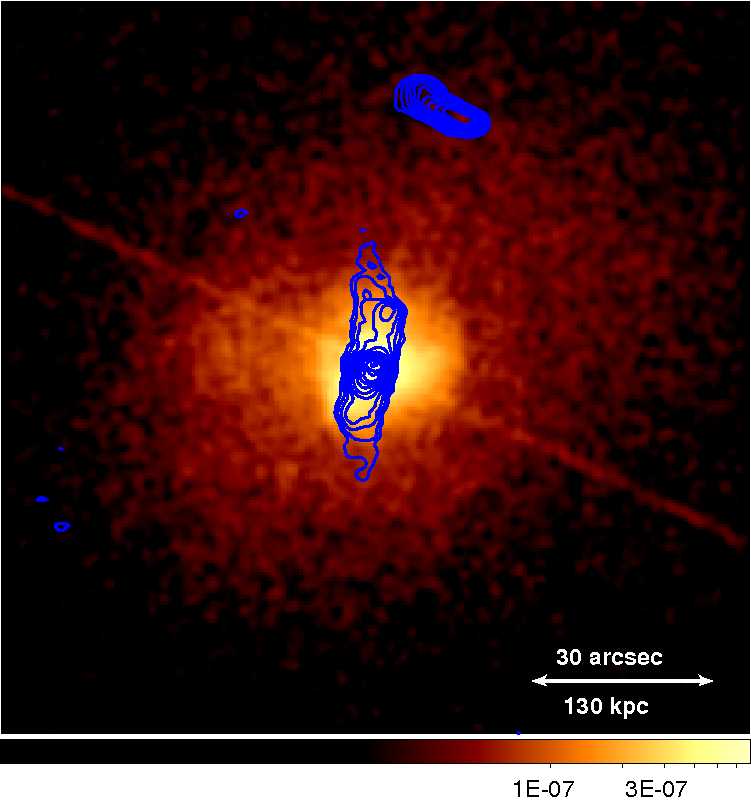}
\caption{Left: Exposure-corrected image in the 0.5--7.0$\keV$ energy band smoothed
    with a 2D Gaussian $\sigma=1.5\arcsec$.  The logarithmic scale bar
    has units $\expmap$.  Right: with superimposed
  VLA $1.4\GHz$ radio contours from \citet{Blundell01}.}
\label{fig:mainimage}
\end{minipage}
\end{figure*}

In Fig. \ref{fig:mainimage} we show a merged exposure-corrected
image of H1821+643 centred on the bright, piled up quasar, with
superimposed $1.4\GHz$ radio contours from \citet{Blundell01}.  The
extended cluster emission surrounding the quasar is elongated in a
north-west to south-east direction.  A surface brightness edge is also
visible around the cluster core at a radius of approximately
$15\arcsec$ from the quasar.  The edge, possibly indicative of a shock
in the cluster gas, is most sharply defined around the north-west of
the cluster core.  

\begin{figure*}
\begin{minipage}[t]{\textwidth}
\centering
\raisebox{0.78 cm}{
\includegraphics[width=0.32\columnwidth]{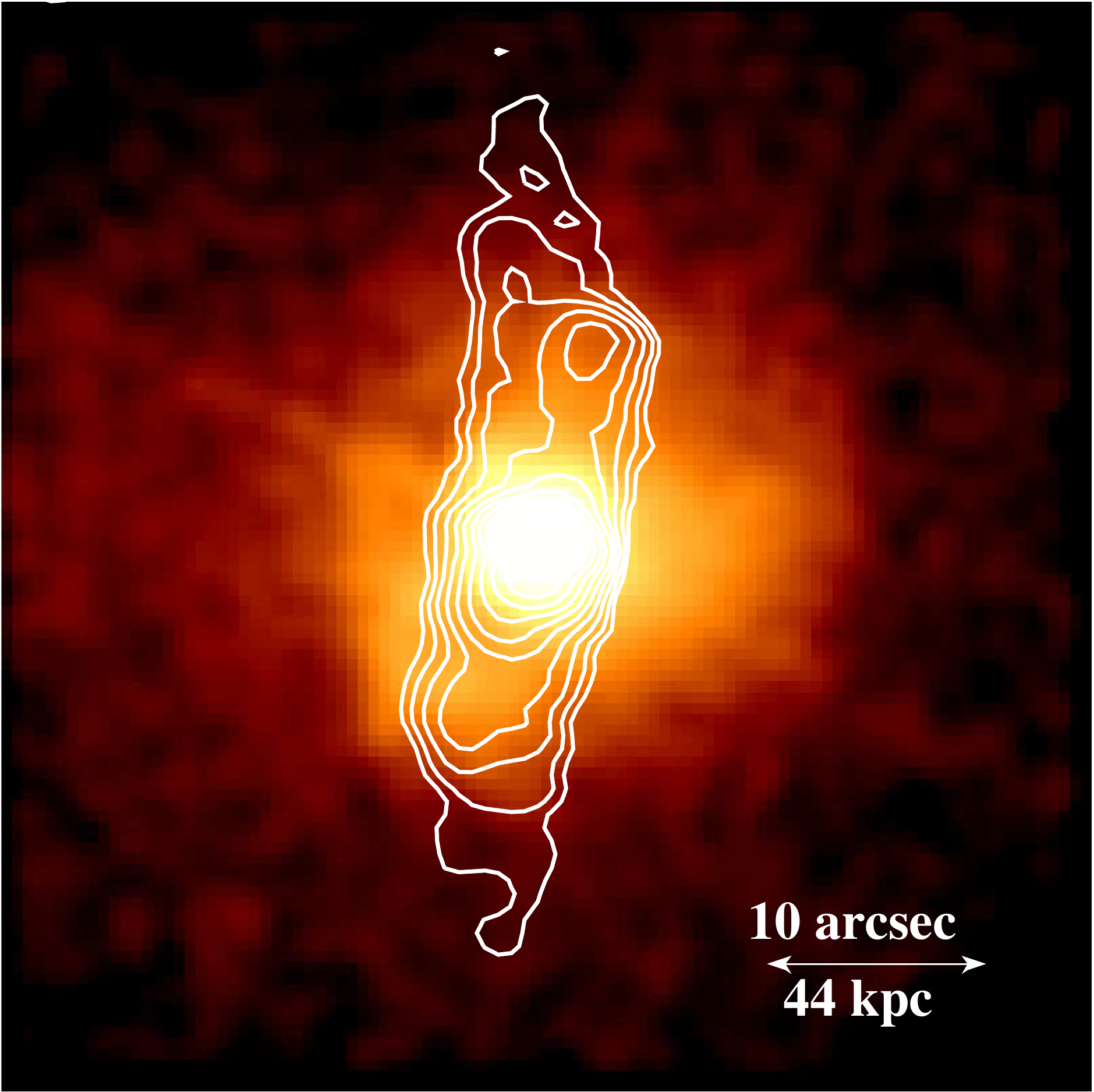}
}
\includegraphics[width=0.32\columnwidth]{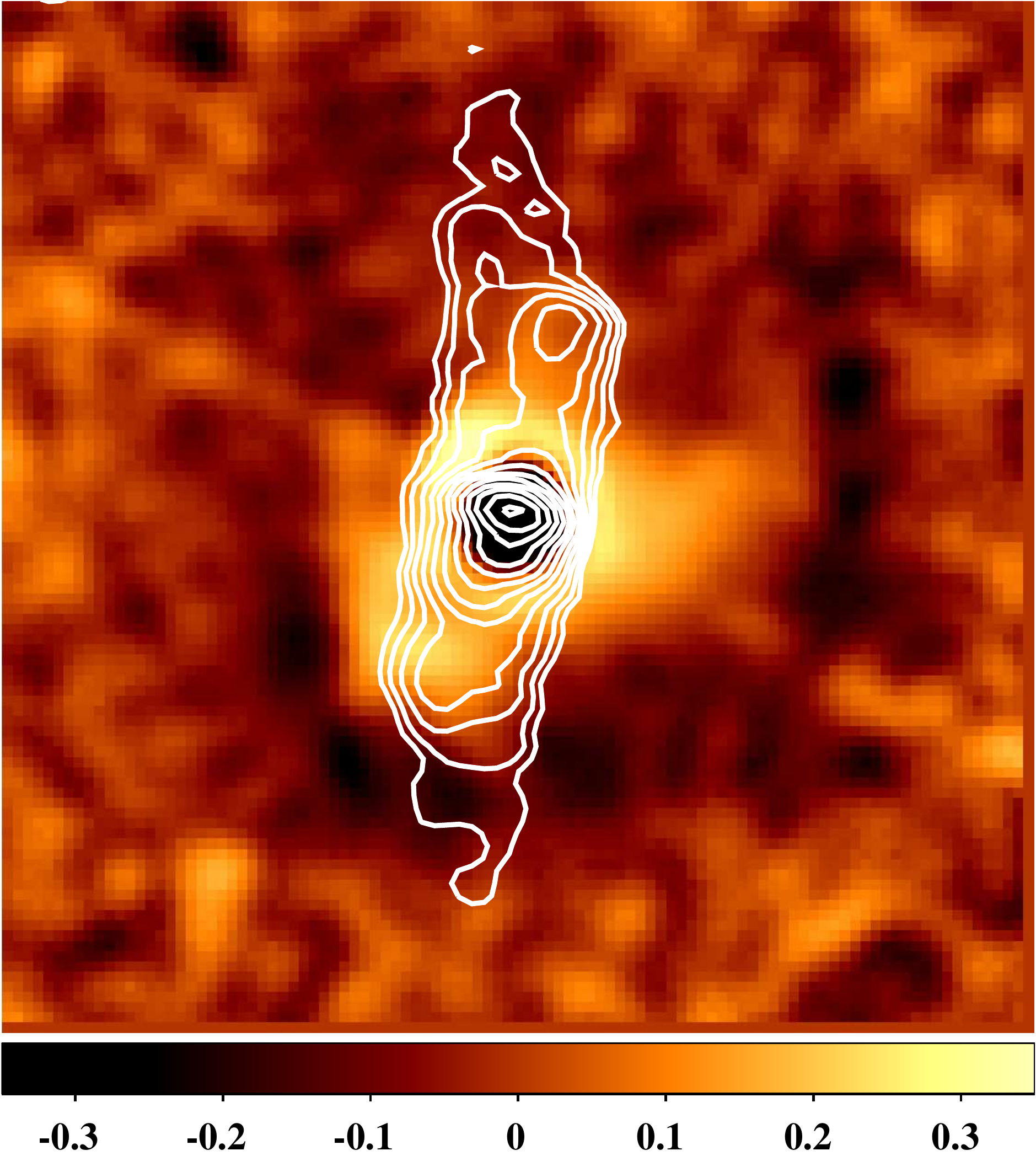}
\includegraphics[width=0.32\columnwidth]{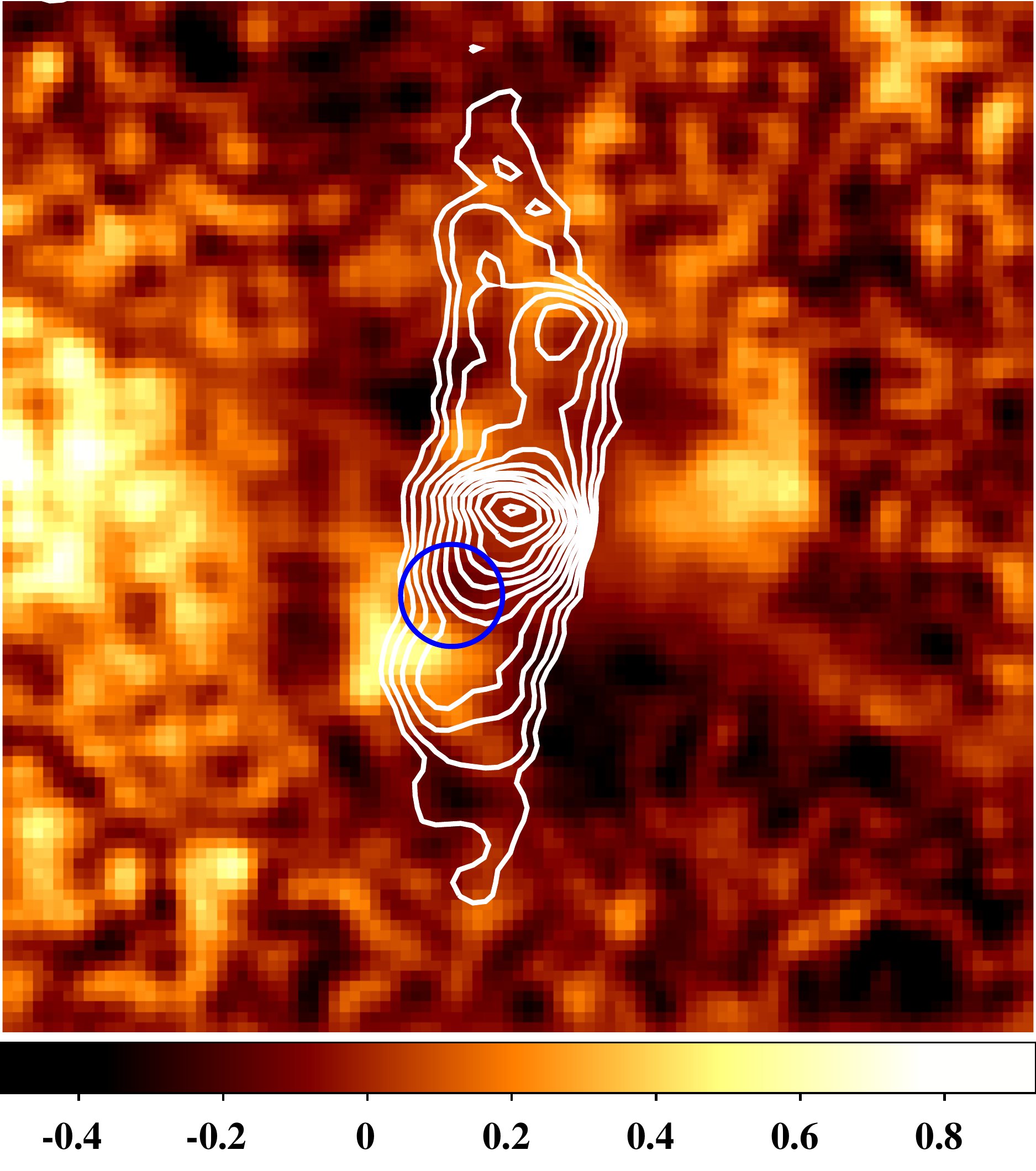}
\caption{Images of the central $0.8\times0.8\arcmin$ with superimposed
  VLA $1.4\GHz$ radio contours from \citet{Blundell01}.  Left:
  Exposure-corrected image in the 0.5--7.0$\keV$ energy band smoothed
  with a 2D Gaussian $\sigma=1.5\arcsec$.  Centre: Unsharp-masked
  image, created by subtracting images smoothed by 2D Gaussians with
  $\sigma=1$ and $5\arcsec$ and dividing by the sum of the two images
  (RMS noise $\sim0.1$).  The central $3\arcsec$ region containing
  the majority of the quasar emission was masked out before smoothing.
  Right: Fractional difference of each pixel from the average at that
  radius (RMS noise $\sim0.2$).  The resulting image has been
  smoothed with a 2D Gaussian $\sigma=1.5\arcsec$.  The SE X-ray
  depression is marked by a blue circle.}
\label{fig:unsharpsub}
\end{minipage}
\end{figure*}

The unsharp-masked image (Fig. \ref{fig:unsharpsub} centre) highlights
the edge in the surface brightness which can be seen to run round the
cluster core from the north-west to south-east.  This does not appear
to be a symmetrical feature about the quasar; the radius of the edge
varies from $13\arcsec$ to $17\arcsec$.  Although the X-ray emission
is extended to the north and south, there does not seem to be a close
association with the $1.4\GHz$ radio emission.  The fractional
difference image (Fig. \ref{fig:unsharpsub} right) shows a bright
extended spur of emission to the south-east of the AGN containing a
surface brightness depression which could indicate a cavity. This
X-ray depression is coincident with the inner part of the radio lobes
but there is no close correlation with the outer region of the radio
emission.  Projection effects may have prevented a clean
interpretation of the relation between X-ray and radio emission.

There are potentially three additional cavities in the X-ray emission
(north-west, north-east and south-west of the quasar, Fig.
\ref{fig:unsharpsub} right).  However, these are not coincident with
any $1.4\GHz$ radio emission and it is not clear how they could have
reached their current locations.

To more quantitatively investigate the interaction between the quasar
and the cluster, we extracted the properties of the cluster gas as
close in to the quasar as possible. 

\section{Spectral Analysis}
\subsection{Separation of Quasar and Cluster}
\label{sec:separation}
Determining the properties of the ICM in the centre of the galaxy
cluster requires a reliable separation of the quasar and cluster
emission.  Although the cluster emission completely dominates over the
quasar emission outside $\sim6\arcsec$ from the centre, we are
particularly interested in probing the gas cooling times as close as
possible to the quasar.  Naively adding an absorbed power-law
model, with freely fitting parameters, to account for this quasar
emission gave a power-law normalization that was an order of magnitude
below the expected value.  The contribution of the quasar should be
comparable to the cluster emission in the innermost radial bins
suggesting that the majority of the quasar spectrum was incorrectly
interpreted in the spectral fitting as hot cluster emission.  We
therefore generated a simulation of the quasar PSF to explicitly
account for the emission spilt-over from the quasar onto the cluster.

The \emph{Chandra} PSF has a complex structure that can be
approximately divided into two sections: a core produced by
quasi-specular X-rays reflecting from the mirror surface and wings
generated by diffracted X-rays scattering from high frequency surface
roughness.  The diffractive mirror scattering also makes the PSF wings
energy dependent: the high energy PSF is broader than at low energy.
A detailed analysis of the \emph{Chandra} PSF produced by the high
resolution mirror assembly (HRMA) can be found on the CXC
website\footnote{See http://cxc.harvard.edu/cal/Hrma/psf/}.

We used the \emph{Chandra} ray-tracing program ChaRT
(\citealt{Carter03}) and the \textsc{marx} software\footnote{See
  http://space.mit.edu/CXC/MARX/} version 4.4, which projects the rays
onto the detector, to generate simulated observations of the quasar.
This allowed an analysis of the spatial and energy dependence of the
PSF without the complication of cluster emission.  ChaRT takes as
inputs the position of the point source on the chip, exposure time and
the quasar spectrum.  However, because the core of the H1821+643
observation is piled up, the quasar spectrum could not be determined
simply from a small, quasar-dominated region on the chip.  Instead,
the quasar spectrum was extracted from the readout streak.

The analysis of the readout streak spectrum, a ChaRT simulation of the
quasar and a validation of this method for the test object 3C\,273 are
detailed in the appendix.  In summary, although ChaRT does not provide
an exact prediction of the observed PSF, the steep decline in the
quasar contribution with radius compared to the cluster emission
mitigated the effect of underestimating the PSF wings beyond
$10\arcsec$.  To test the accuracy of the ChaRT simulations, we
repeated our method for a Chandra observation of 3C\,273 (obs
ID. 4879), excluding the region containing the jet.  The simulation
accurately predicted the observed 3C\,273 quasar spectra that were
extracted in annuli centred on the nucleus.  We therefore proceeded
with the ChaRT simulation of H1821+643 and analysed the uncertainties
introduced by this method in section \ref{sec:SBresults}.

\subsection{Projected Radial Profiles}
\label{sec:projmethod}
\begin{figure*}
\begin{minipage}[t]{\textwidth}
\centering
\includegraphics[width=0.48\columnwidth]{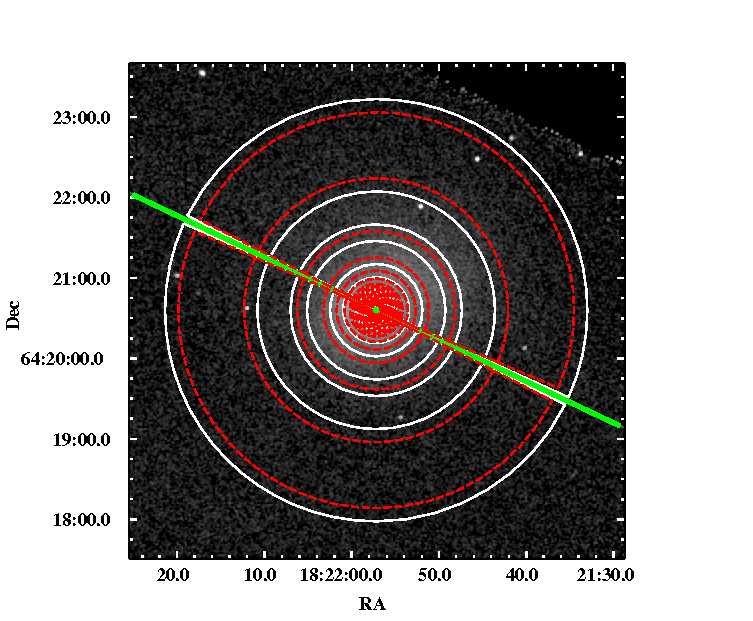}
\includegraphics[width=0.48\columnwidth]{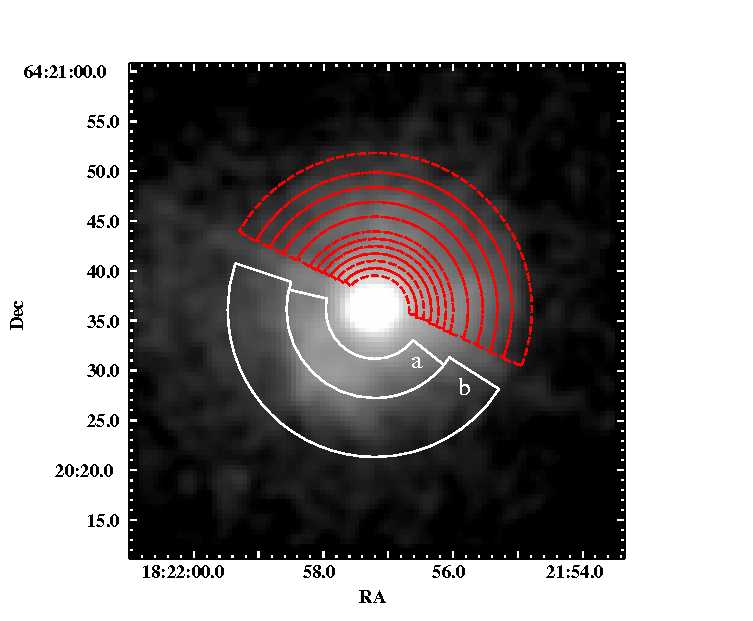}
\caption{Exposure-corrected images in the 0.5--7.0$\keV$ energy band
  smoothed with a 2D Gaussian $\sigma=1.5\arcsec$.  Left: The regions for
  projected spectra are shown by the magenta dashed lines (referred to as 1-19
  from the centre outwards) and those for deprojected spectra are
  shown by the white solid lines (referred to as a-i from the centre
  outwards).  The readout streak, shown by the green box, and point
  sources were excluded. Right: Zoom in of central region showing the
  annuli affected by quasar emission (only half of each annulus is
  shown for clarity).}
\label{fig:obsregions}
\end{minipage}
\end{figure*}

Radial profiles of gas properties, such as temperature and density,
were generated to analyse the gas properties in the core of the galaxy
cluster.  Spectra were extracted in a series of concentric annuli
centred on the emission peak and excluding the readout streak and
point sources (Fig. \ref{fig:obsregions}).  The radial bins were
chosen to ensure a minimum of 3000 counts in each extracted spectrum
with the minimum radius determined from the pileup analysis to be
$3\arcsec$.  This criterion ensured enough counts in each spectrum to
provide a good spectral fit and constraints on the cluster properties.
Point sources were identified using the CIAO algorithm
\textsc{wavdetect}, visually confirmed and excluded from the analysis
using elliptical apertures where the radii were conservatively set to five times the
measured width of the PSF.  All spectra were analysed in the energy
range $0.5-7\keV$ and grouped with a minimum of 50 counts per spectral
bin.  The background was subtracted using a spectrum extracted in a
sector at large radii, $170-210\arcsec$.  Response and ancillary
response files were generated for each cluster spectrum, weighted
according to the number of counts between 0.5 and $7\keV$.

These projected spectra were then fitted in \textsc{xspec} version
12.5.0 (\citealt{Arnaud96}) with an absorbed power-law model, to
account for the quasar PSF, and an absorbed thermal plasma emission
model \textsc{phabs(powerlaw) + phabs(mekal)} (\citealt{Balucinska92};
\citealt{Mewe85}; \citeyear{Mewe86}; \citealt{Kaastra92};
\citealt{Liedahl95}).  Abundances were measured assuming the abundance
ratios of \citet{AndersGrevesse89}.  The parameters describing the
quasar absorbed power-law model were fixed to the values given in
Table \ref{powvalues}.  These values were determined by fitting a
\textsc{phabs(powerlaw)} model to spectra extracted from the
appropriate regions in the ChaRT simulation.  The variation in the
values of these parameters with radius is not caused by an intrinsic
variation in the emission from the source.  Instead, this variation is
produced by the difference in the effective area of the detector for
a small fraction of the quasar PSF compared to the cluster emission
(see appendix).

For the cluster model component, the temperature, abundance and normalization
parameters were all left free.  The absorbing column density was fixed
to $n_{\mathrm{H}}=4\times10^{20}\pcmsq$ determined from spectral
fitting in the outer annuli.  This value was consistent within error
with the Galactic value $n_{\mathrm{H}}=3.44\times10^{20}\pcmsq$
(\citealt{Kalberla05}).  The redshift was fixed to $z=0.299$
(\citealt{Schneider92}).  

\begin{table*}
\begin{minipage}[t]{\textwidth}
\centering
\caption{Table of best-fitting quasar parameters for each region.  Column
  1) Region name 2) Inner and outer radii of annulus (\arcsec) 3)
  Galactic column density (\pcmsq) 4) Photon index 5) \textsc{xspec}
  power-law normalization ($\plawnorm$ at 1\keV) 6) $\chi^{2}$ /
  number of degrees of freedom}
\setlength{\extrarowheight}{3pt}
\begin{tabular}{l c c c c c}
\hline 
Region & Radius & n$_{H}$ & Photon index & Norm & $\chi^{2}$/dof \\
 &  & $10^{22}$ & & & \\
\hline
proj$_{1}$ & 3.4--4.2 & $0.06\pm0.01$ & $1.59\pm0.03$ &
$1.63\pm0.08\times10^{-5}$ & 72/78 \\
proj$_{2}$ & 4.2--4.9 & $0.03\pm0.01$ & $1.52\pm0.03$ &
$1.14^{+0.07}_{-0.06}\times10^{-5}$ & 64/59 \\
proj$_{3}$ & 4.9--5.7 & $0.04\pm0.01$ & $1.52^{0.04+}_{-0.05}$ &
$8.8\pm0.6\times10^{-6}$ & 55/44 \\
proj$_{4}$ & 5.7--6.4 & $0.04\pm0.01$ & $1.50\pm0.04$ &
$7.1^{+0.5}_{-0.5}\times10^{-6}$ & 33/36 \\
proj$_{5}$ & 6.4--7.1 & $0.09\pm0.02$ & $1.60\pm0.05$ &
$6.5^{+0.6}_{-0.5}\times10^{-6}$ & 25/29 \\
proj$_{6}$ & 7.1--7.9 & $0.004^{+0.02}_{-0.004}$ & $1.47^{+0.05}_{-0.06}$ &
$4.5^{+0.5}_{-0.2}\times10^{-6}$ & 25/24 \\
proj$_{7}$ & 7.9--9.3 & $0.01\pm0.01$ & $1.45^{+0.04}_{-0.07}$ &
$7.2^{+0.6}_{-0.4}\times10^{-6}$ & 31/38 \\
proj$_{8}$ & 9.3--10.8 & $0.01\pm0.01$ & $1.43^{+0.04}_{-0.07}$ &
$5.7^{+0.5}_{-0.3}\times10^{-6}$ & 28/32 \\
proj$_{9}$ & 10.8--12.3 & $<0.01$ & $1.37^{+0.06}_{-0.04}$ &
$4.1^{+0.3}_{-0.1}\times10^{-6}$ & 23/23 \\
proj$_{10}$ & 12.3--13.8 & $0.01^{+0.02}_{-0.01}$ & $1.33^{+0.06}_{-0.1}$ &
$3.3^{+0.4}_{-0.2}\times10^{-6}$ & 8/17 \\
proj$_{11}$ & 13.8--15.7 & $0.03\pm0.02$ & $1.48^{+0.05}_{-0.06}$ &
$4.1\pm0.4\times10^{-6}$ & 14/21 \\
\\
deproj$_{a}$ & 5--9 & $0.06\pm0.01$ & $1.57\pm0.05$ &
$3.8\pm0.2\times10^{-3}$ & 101/95 \\
deproj$_{b}$ & 9--15 & $0.006^{+0.01}_{-0.006}$ & $1.41^{+0.05}_{-0.04}$ &
$5.8^{+0.3}_{-0.2}\times10^{-4}$ & 84/95 \\
\hline
\end{tabular}
\label{powvalues}
\end{minipage}
\end{table*}

\subsection{Deprojected Radial Profiles}
\label{deprojection}
A spectrum extracted from the centre of the cluster on the plane of
the sky corresponds to a summed cross-section with a range of spectral
components from the core to the cluster outskirts.  To determine the
properties of the cluster core these projected contributions from the
outer cluster layers can be subtracted off the inner spectra by making
an assumption about the line of sight extent and deprojecting the
emission.

In this work we used two deprojection routines: a straightforward
spectral deprojection (\textsc{dsdeproj}; \citealt{SandersFabian07};
\citealt{Russell08}) assuming only spherical symmetry and an updated
version of the `X-ray surface brightness deprojection' code, which
uses the additional assumptions of hydrostatic equilibrium and a
suitable mass model, and was first described in \citet{Fabian81} (see
also \citealt{White97}; \citealt{Allen01}; \citealt{Schmdit01}).

\begin{figure}
\centering
\includegraphics[width=\columnwidth]{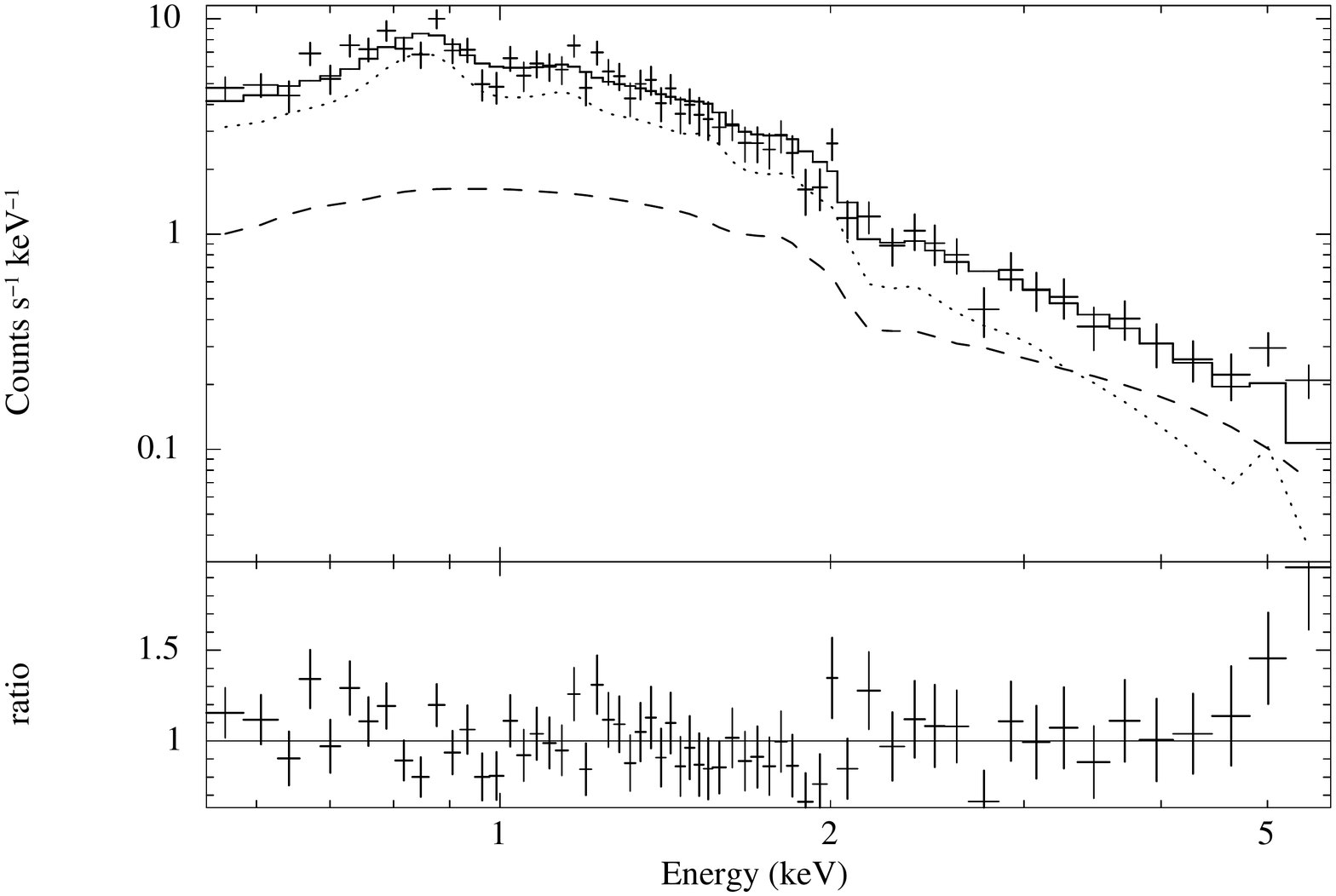}
\includegraphics[width=\columnwidth]{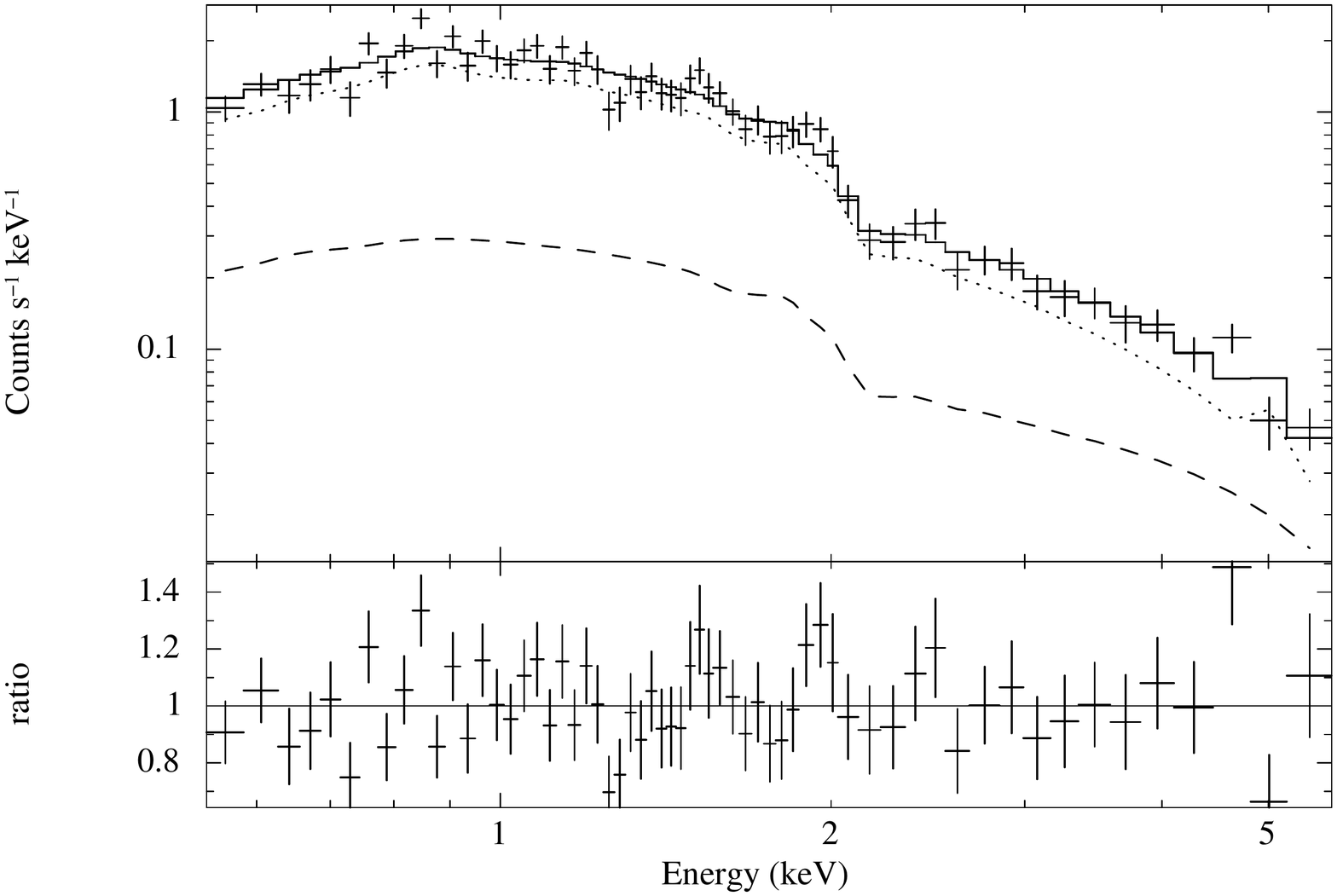}
\caption{Spectrum and best-fitting model (solid line) for the inner
  \textsc{dsdeproj} deprojected
  spectra requiring a quasar emission model: deproj$_a$,
  $5-9\arcsec$ (upper), and deproj$_{b}$, $9-15\arcsec$
  (lower).  The model is a combination of a power-law model (dashed
  line) to account for the quasar emission and thermal plasma model
  (dotted line) for the cluster emission.}
\label{fig:innerspectrafits}
\end{figure}

In more detail, \textsc{dsdeproj} starts with the
background-subtracted spectrum extracted from the outermost annulus
and assumes it was emitted from part of a spherical shell.  This
spectrum is scaled by the volume projected onto the next innermost
shell (geometric factors from \citealt{Kriss83}) and subtracted from
the spectrum of that annulus.  In this way the deprojection routine
moves inwards subtracting the contribution of projected spectra from
each successive annulus to produce a set of deprojected spectra.  The
input projected spectra were analysed in the energy range $0.5-7\keV$
and grouped with a minimum of 50 counts per spectral bin as before.
The radial bins were chosen to ensure a minimum of 3000 counts in each
deprojected spectrum and so are larger than the radial bins for the
projected spectra.  The resulting deprojected spectra were also fitted
with an absorbed power-law model to account for the quasar PSF and an
absorbed thermal plasma emission model.  The parameters describing the
quasar model were fixed to the values determined from the ChaRT
simulation, where the ChaRT spectra were also deprojected (Table
\ref{powvalues}).  The cluster spectral model parameters were set as
detailed in section \ref{sec:projmethod}.  The spectra and
best-fitting models for the inner deprojected spectra with significant
contributions from the quasar emission are shown in Fig.
\ref{fig:innerspectrafits}.

The X-ray surface brightness deprojection code uses the observed
quasar-subtracted surface brightness profile, an NFW mass model
(\citealt{NFW97}) and assumes hydrostatic equilibrium to predict the
temperature profile of the gas.  The predicted temperature profile is
then rebinned to match the \textsc{dsdeproj} result and the $\chi^2$
difference between the observed and predicted deprojected temperature
profiles is calculated.  The mass model parameters of concentration
and scale radius are iterated over using a Markov Chain Monte Carlo
method to determine the best-fitting values which give the minimum
$\chi^2$.  The MCMC routine was constructed with a Metropolis-Hastings
algorithm comparing four chains of length 190,000 with a burn-in length of
10,000.  The metallicity was fixed to $0.4\Zsun$ for this analysis.

This analysis assumes that the detected X-rays from this source
were produced either by the quasar nucleus or by bremsstrahlung
and line emission from the ICM.  However, there could be a
contribution from X-ray synchrotron or inverse-Compton emission
produced by the radio jets and lobes.  The $1.7\GHz$ MERLIN
observation of H1821+643 showed that the radio jets were sited within
a radius of $\sim2\arcsec$, therefore an analysis of the X-ray emission
at radii above $3\arcsec$ should not contain significant X-ray synchrotron
radiation.  The contribution of inverse-Compton emission, generated by
relativistic plasma in the radio source lobes upscattering the
jet-produced photons or the cosmic microwave background, was difficult
to estimate.  However, the spectra from $3\arcsec$ were well-described
by the two component spectral fits incorporating
the quasar and cluster emission and there is unlikely to be a
significant extra contribution from inverse-Compton emission.

\section{Results}
\subsection{Cluster surface brightness profile}
\label{sec:SBresults}
\begin{figure}
\centering
\includegraphics[width=0.9\columnwidth]{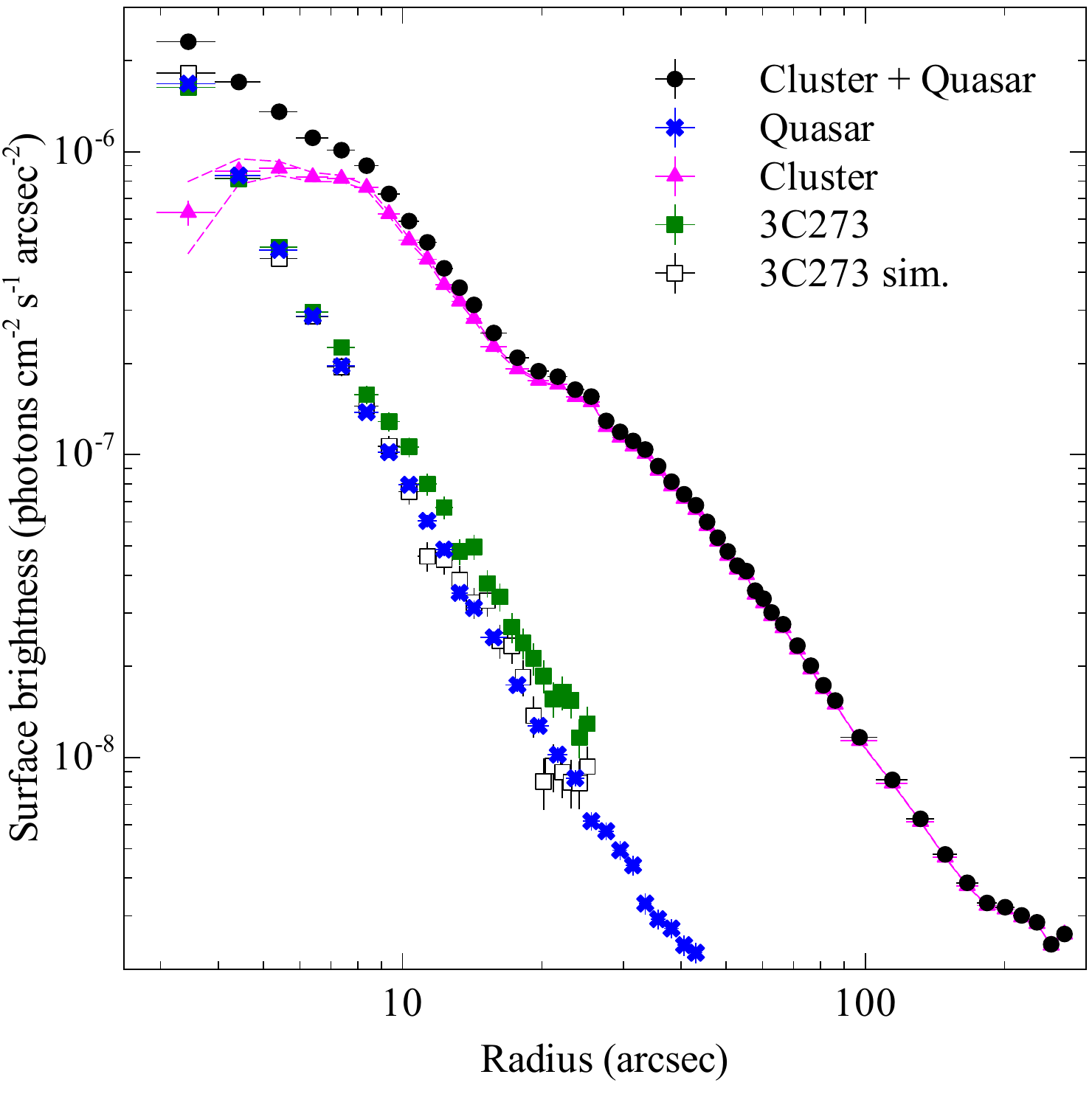}
\caption{Total surface brightness profile in the energy range
  $0.5-7.0\keV$ for H1821+643 (black circles), ChaRT simulation of the
  quasar surface brightness profile (blue crosses) and the cluster
  surface brightness profile produced by subtracting the quasar
  simulation from the total (magenta triangles).  The surface brightness
  profile for 3C\,273 (green squares) and corresponding simulation
  (white squares) are shown for comparison.  The 3C\,273 profiles have
  been normalized by the ratio of the $0.5-7.0\keV$ fluxes as
  determined from the readout streaks.  The magenta dashed lines show the
  effect of increasing and decreasing the quasar contribution by 10\%.}
\label{fig:SBprofile}
\end{figure}

The total surface brightness profile calculated from the
exposure-corrected image for H1821+643 is shown in Fig.
\ref{fig:SBprofile}.  The radial bins are 1\arcsec wide in the
cluster centre and this was increased at larger radii where the background
becomes more important.  The simulated quasar PSF was subtracted from the
total surface brightness profile to produce the cluster surface
brightness profile.  The flattening of the cluster profile from
$15-20\arcsec$ corresponds to the surface brightness edge seen in
Fig. \ref{fig:mainimage}.  In addition, a second break in the
cluster surface brightness profile can be seen at $\sim8\arcsec$.  The
cluster surface brightness profile decreases in the innermost radial
bin suggesting that the quasar component could have been
overestimated.  

Therefore, before proceeding with any analysis of these features, it
was important to determine the uncertainty in the cluster surface
brightness introduced by the quasar subtraction.  This was achieved by
comparing the observation and ChaRT simulation of 3C\,273 as an
estimate for the accuracy of the simulation of H1821+643.  Fig.
\ref{fig:SBprofile} shows a comparison of the ChaRT simulation of
H1821+643 with the observation and simulation of 3C\,273.  The surface
brightness profiles for 3C\,273 were scaled by the ratio of the fluxes
of the two quasars, determined from the respective readout streak
spectra in the $0.5-7.0\keV$ energy band.  The ChaRT simulation was
not expected to provide an exact prediction of the observed PSF, and
indeed for 3C\,273 the simulation underpredicts the observation,
particularly in the PSF wings outside $10\arcsec$.  The ChaRT model
for the PSF wings is known to underestimate the wings\footnote{See
  http://cxc.harvard.edu/chart/caveats.html} and therefore this
underestimation is likely to also be a problem for H1821+643.  

However in H1821+643 the underprediction outside $10\arcsec$ is
mitigated by the steep decline in the quasar contribution with radius
compared to the cluster emission.  For the region inside $10\arcsec$
where the subtraction is important, the simulation of 3C\,273
underpredicts the observation by up to $\sim10\%$.  This offset could
be due to the ChaRT model or alternatively caused by an
underestimation in the modelling of the readout streak.  Therefore, we
estimated that the simulated quasar profile for H1821+643 could be
offset by $\pm10\%$ inside $10\arcsec$ radius.  This offset did not
significantly affect the quasar spectral parameters determined from
the simulated spectra extracted in large radial bins, which were used
to determine the cluster properties close to the nucleus
(Fig. \ref{fig:3C273compare}).  However, for the cluster surface
brightness profile, the offset resulted in an uncertainty of 25\% in
the innermost cluster profile radial bin, dropping to 5\% by the
third bin and then an insignificant contribution above $\sim6\arcsec$
(Fig.  \ref{fig:SBprofile}).  Therefore the decrease in cluster
surface brightness observed in the central radial bin was not
significant given the dominating contribution of the quasar.

\begin{figure}
\centering
\includegraphics[width=0.8\columnwidth]{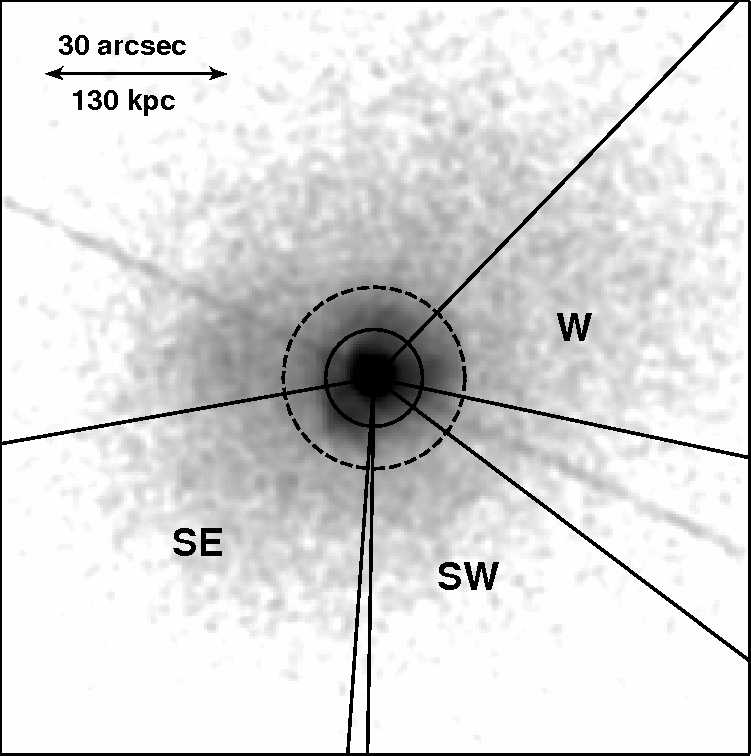}
\caption{Exposure-corrected image in the 0.5--7.0$\keV$ energy band smoothed
    with a 2D Gaussian $\sigma=1.5\arcsec$.  The labelled sectors
    correspond to the area used to produce the surface brightness
    profiles in Fig. \ref{fig:deprojSBprofile}.  The two surface
    brightness edges are marked with circles at $8\arcsec$ (solid) and
    $15\arcsec$ (dashed).}
\label{fig:SBsectors}
\end{figure}

The cluster surface brightness profile was also deprojected and
evaluated in sectors (Fig. \ref{fig:SBsectors}) to study the core
structures.  The width of the radial bins was increased from those of
projected radial profile to ensure enough deprojected counts in each
bin and produce a reliable deprojection.  The quasar contribution was
first subtracted from the total surface brightness profile.  Then the
resulting cluster surface brightness profile was deprojected by
assuming spherical symmetry to subtract the contribution from the
outer annuli off of the inner annuli.  The region from $3-5\arcsec$
was excluded from the deprojection because the shallow gradient of the
core surface brightness profile produced large errors in the values,
in addition to the uncertainty from the quasar subtraction.

The SW sector (magenta) contains a section of the surface brightness
edge at $\sim15\arcsec$ but no extended emission from the core.
Fig. \ref{fig:unsharpsub} (right) suggests there could be a `ghost' cavity
from $5-12\arcsec$ in this sector.  By analysing VLA $1.4\GHz$ and
MERLIN $1.7\GHz$ observations, \citet{Blundell01} concluded that the jet axis in
H1821+643 could be precessing.  The `tipped over' appearance of the N plume of radio
emission in Fig. \ref{fig:mainimage} is suggestive of this.  The X-ray depressions
to the NE and SW identified in Fig. \ref{fig:unsharpsub} could have been inflated by the
jet at an earlier time, detached and risen buoyantly as the jet
precessed.

However the deprojected surface brightness profile does not appear to
support this interpretation.  Although there is less emission from
this sector, the surface brightness profile inside $14\arcsec$ is
smooth without the clear decrement expected from a cavity.  It is
therefore likely that the fractional difference image (Fig.
\ref{fig:unsharpsub} right) highlights the excess emission in the
extended arms compared to the remaining emission at the same radius.
However, cluster substructure and deviations from spherical symmetry
in the cluster core will cause problems for the deprojection routine,
particularly at radii inside the surface brightness edge.  We cannot
rule out the possibility that these features are `ghost' cavities
(X-ray cavities without any detected coincident radio emission).

\begin{figure}
\centering
\includegraphics[width=0.95\columnwidth]{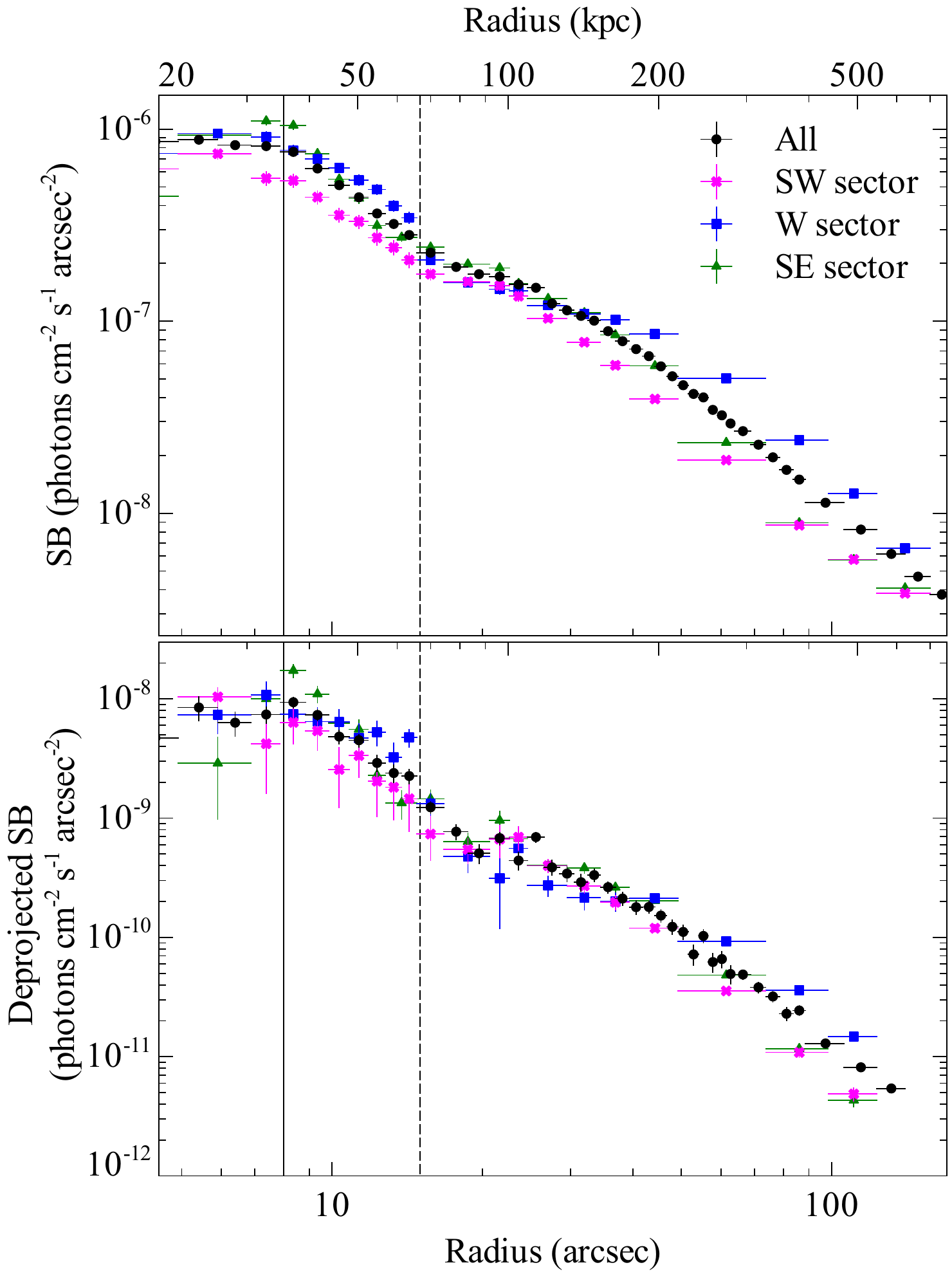}
\caption{Top: quasar-subtracted surface brightness profiles in the energy range
  $0.5-7.0\keV$ for H1821+643 extracted in sectors (Fig.
  \ref{fig:SBsectors}) and the total surface brightness profile shown
  for comparison (black circles).  Bottom: quasar-subtracted deprojected surface
  brightness profiles.  The solid and dashed lines correspond to the
  circles in Fig. \ref{fig:SBsectors}.}
\label{fig:deprojSBprofile}
\end{figure}

The SE (green) and W (blue) sectors both contain extended arms of
emission from the core and for the W sector this terminates
particularly abruptly at the surface brightness edge at $15\arcsec$
(dashed line, Fig. \ref{fig:SBsectors} and
Fig. \ref{fig:deprojSBprofile}).  For these two sectors there is also
evidence for an inner surface brightness edge at $8\arcsec$ (solid line,
Fig. \ref{fig:SBsectors} and Fig. \ref{fig:deprojSBprofile}).  Behind
this inner edge, there could be a radio cavity in the SE sector.
However, although the surface brightness profile is consistent with a
decrease at this radius, there are only two radial bins in this region
and they contain additional uncertainty associated with the quasar
subtraction.  Therefore, it is not clear whether this is a radio
cavity.



\subsection{Projected Spectral Analysis}
\begin{figure}
\centering
\includegraphics[width=0.95\columnwidth]{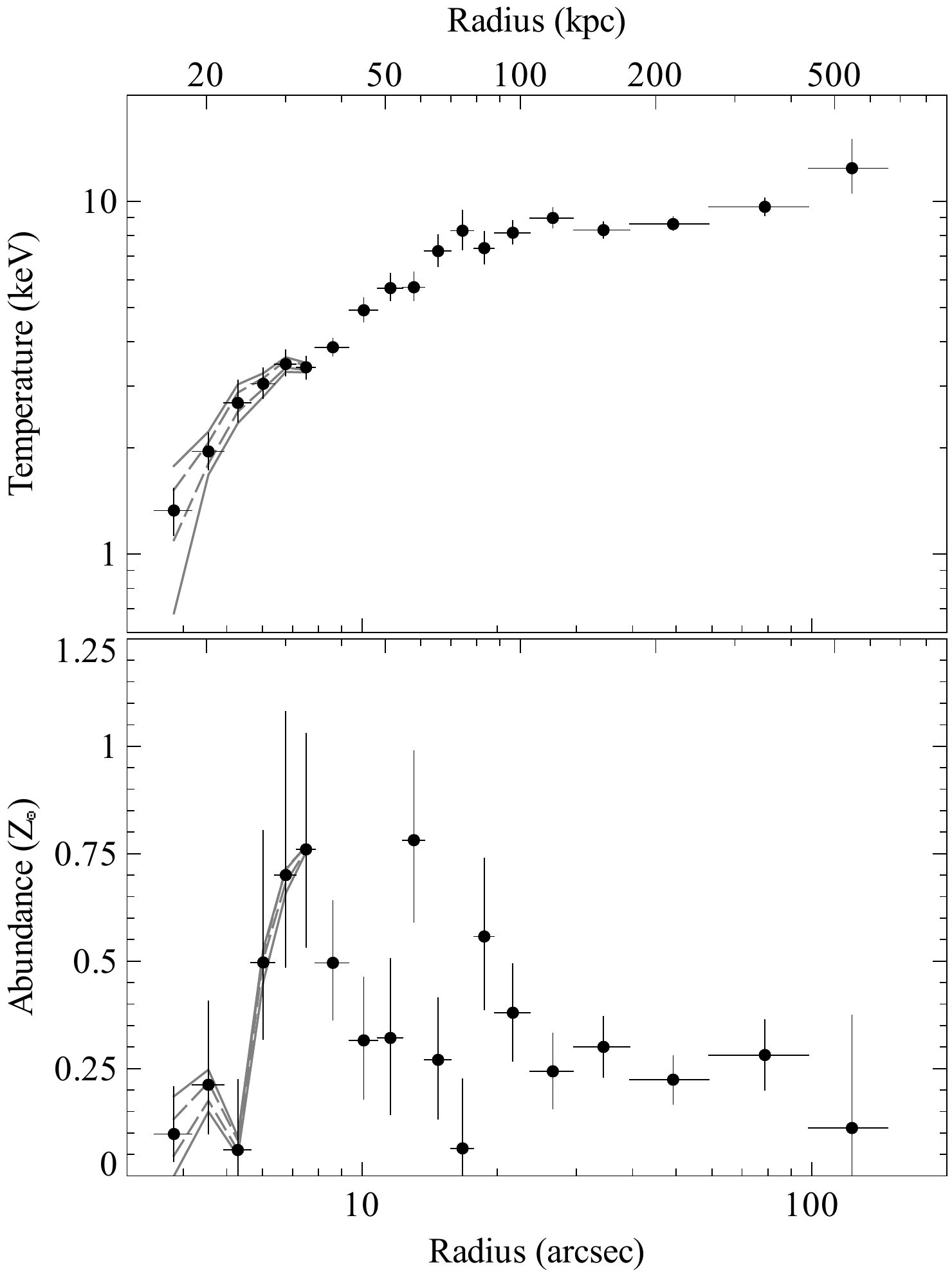}
\caption{Projected radial temperature (upper) and metallicity (lower)
  profiles.  The quasar contribution has been subtracted.  The
  superimposed lines show the effect of varying the normalization of
  the quasar component by $\pm10\%$ (dashed line) and $\pm20\%$ (solid
  line).}
\label{fig:H1821proj}
\end{figure}

Fig. \ref{fig:H1821proj} shows the projected radial temperature and
metallicity profiles for H1821+643 (also Table
\ref{tab:allprojvalues}).  The temperature profile shows a cool-core
with a broad range in temperature, almost a factor of ten from the
cluster outskirts down to $1.3\pm0.2\keV$ in the centre.  The
temperature starts to decline steeply at $\sim18\arcsec$ and then
breaks again at $\sim7\arcsec$ for a sharp drop into the cluster
centre.  The metallicity is approximately constant at $\sim0.3\Zsun$
at large radii but increases in the cluster core to around
$\sim0.5\Zsun$.  The sudden drop in metallicity seen inside $6\arcsec$
is caused by an underestimate of the Fe\,K\,$\alpha$ line blend at
$\sim5\keV$ (Figure \ref{fig:innerspectrafits}).  This is discussed in
section \ref{sec:quasarimpact}.

We examined the systematic error in the projected radial profiles that
could be introduced if the quasar normalization was varied by
$\pm10\%$ and $\pm20\%$ and the power-law index varied by $\pm0.05$.
Fig. \ref{fig:H1821proj} shows that within a radius of $5\arcsec$ a
variation of $\pm20\%$ in normalization of the quasar component would
add an uncertainty of $\pm0.5\keV$ to the cluster temperature value
and $\pm0.1\Zsun$ to the metallicity.  However, beyond $5\arcsec$ the
contribution of the quasar PSF declines rapidly.  Therefore this
additional error is important only for the central two radial bins.
Variation in the quasar power-law index did not produce significant
additional error in the cluster parameters.

\begin{figure*}
\begin{minipage}[t]{\textwidth}
\centering
\includegraphics[width=0.32\columnwidth]{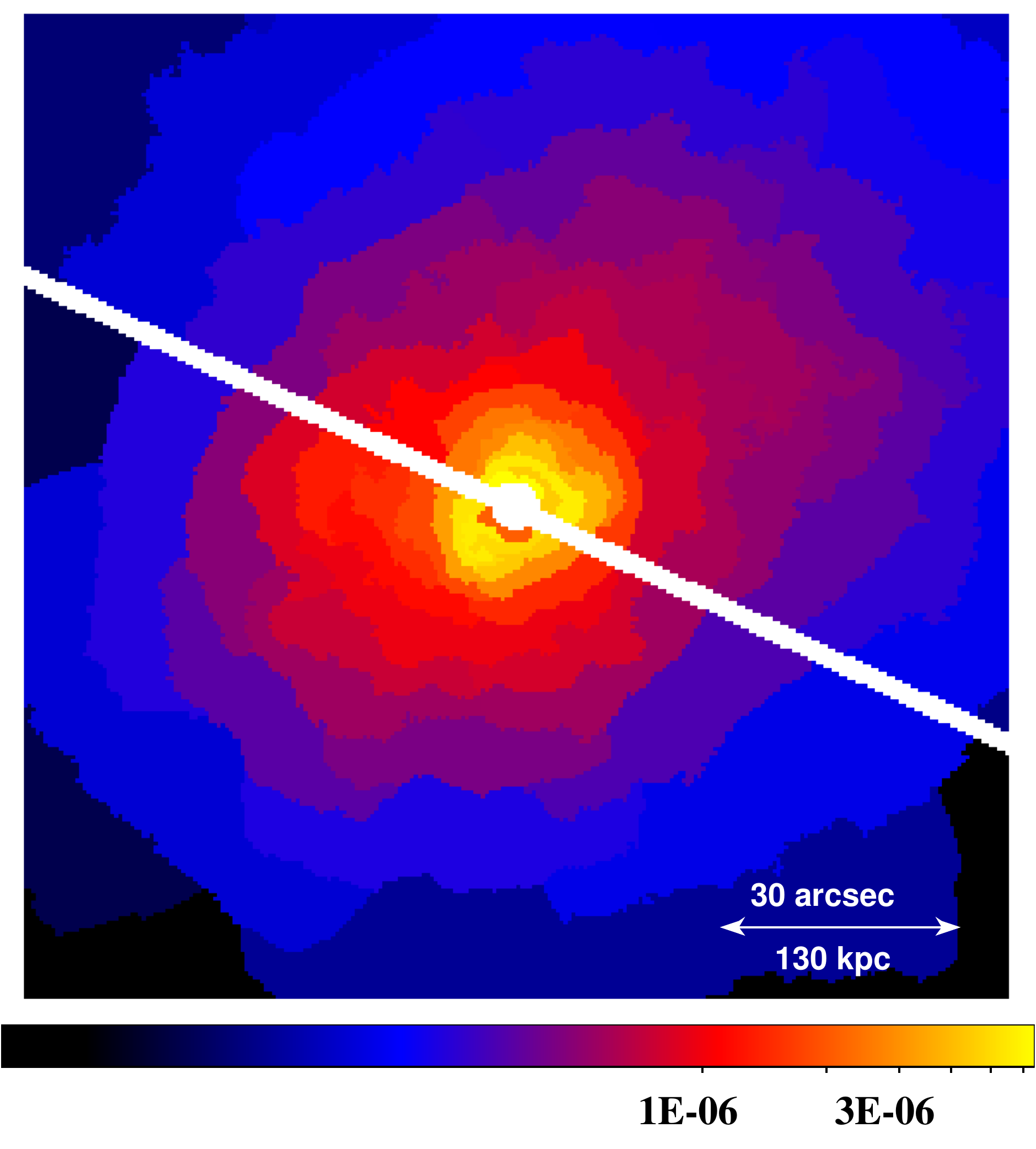}
\includegraphics[width=0.32\columnwidth]{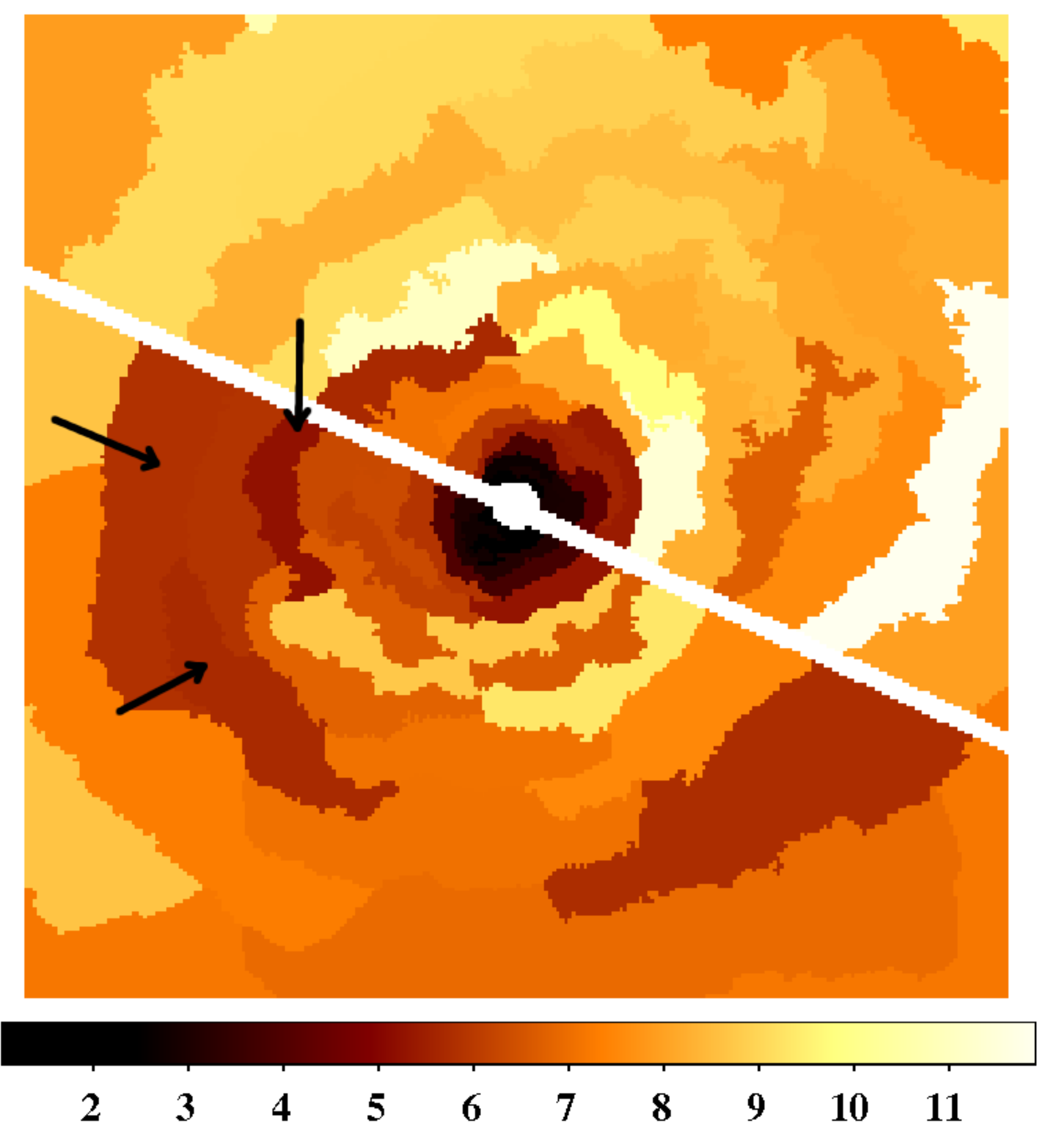}
\includegraphics[width=0.32\columnwidth]{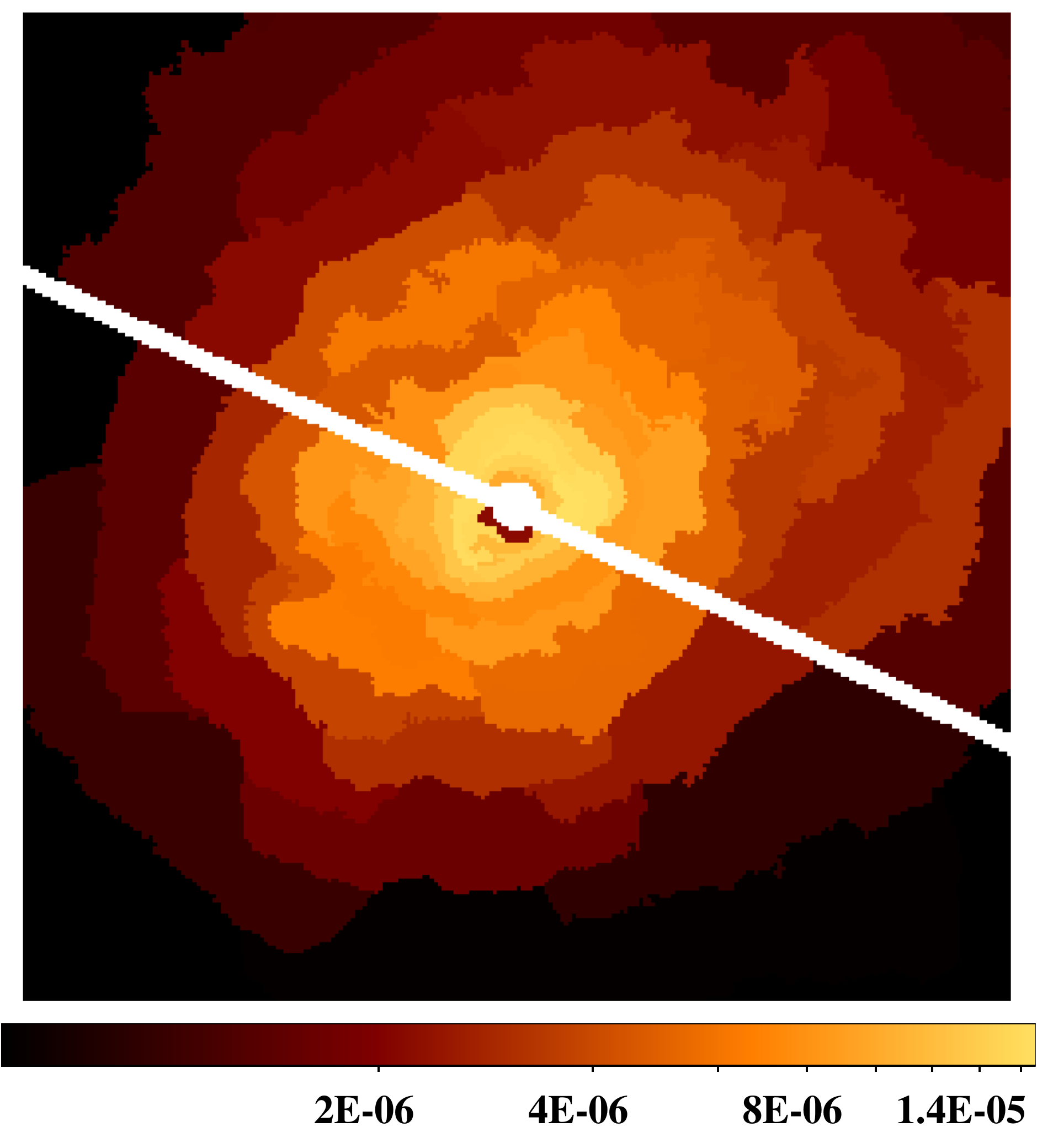}
\caption{Left: emission measure per unit area map (units
  $\empasecsq$).  The emission measure is the \textsc{xspec}
  normalization of the \textsc{mekal} spectrum
  $K=EI/(4\times10^{14}\pi D_A^2(1+z)^2)$, where EI is the emission
  integral $EI=\int n_en_H\mathrm{d}V$.  Centre:
  temperature map ($\keV$) with arrows marking the possible swirl of
  cool gas.  Right: `pressure' map ($\keVempasecsq$)
  produced by multiplying the emission measure and temperature maps.
  The maps cover the same spatial scale and the quasar contribution
  has been subtracted.  The readout streak and
  central $3\arcsec$ radius region affected by pileup have been masked
  out.}
\label{fig:jssmaps}
\end{minipage}
\end{figure*}

We applied spatially resolved spectroscopy techniques to produce maps
of the projected temperature and pressure in the cluster core (Fig.
\ref{fig:jssmaps}).  The central $\sim4\times4\arcmin$ was divided
into bins using the Contour Binning algorithm (\citealt{Sanders06}),
which follows surface brightness variations.  Regions with a
signal-to-noise ratio of 32 ($\sim1000$ counts) were chosen, with the
restriction that the length of the bins was at most two times their
width.  Background and quasar spectra from the ChaRT simulation were
subtracted from the observed dataset and appropriate responses and
ancillary responses were generated.  Each spectrum was fitted with an
absorbed \textsc{mekal} model with the absorption fixed to the
Galactic value and the metallicity fixed to $0.4\Zsun$.  The spectra
were grouped to contain a minimum of 20 counts per spectral channel and
fitted in the energy range $0.5-7\keV$.  The fitting procedure
minimised the $\chisq$-statistic.  The errors were approximately
$\sim10\%$ in temperature and $\sim5\%$ for the emission measure.

Shown in Fig. \ref{fig:jssmaps} are the emission measure per unit
area, temperature and `pressure' maps.  The `pressure' map was
produced by multiplying the emission measure per unit area and the
temperature maps.  The emission measure map shows the strongly peaked
core surface brightness, which flattens in the very centre and the
overall elongated morphology of the cluster in the NW to SE direction.
The temperature in the cluster core drops down to $2\keV$ and there is
a region of cooler gas to the E forming part of a swirl, which could
be similar to the spiral patterns seen in Abell 2029
(\citealt{Clarke04}), Abell 2204 (\citealt{A2204Sanders05}) and in a
temperature map of Perseus (\citealt{FabianPer06}).

The temperature map also shows an increase to the NW from
$15-20\arcsec$ coincident with the strongest section of the surface
brightness edge (Fig. \ref{fig:deprojSBprofile}).  This temperature
increase is also found in a projected temperature profile extracted
in the NW sector (Fig. \ref{fig:SBsectors}), however this increase
is not statistically significant enough to draw any further
conclusions on the nature of the surface brightness edge.  In
addition, the pressure map does not show a particularly sharp drop
from $15-20\arcsec$ which would indicate a shock.  

It is not obvious from any of the maps shown in Fig. \ref{fig:jssmaps} in
which direction the main radio axis is oriented.  Apart from a small
low pressure region to the SE, which could relate to the small radio
bubble, the pressure map appears symmetric about the central source
with no indication that it contains a large FR I source.

\begin{table}
\centering
\caption{Table of best-fitting cluster parameters for each projected region.
  Column 1) Region name 2) Inner and outer radii of annulus (\arcsec)
  3) Temperature ($\keV$) 4) Abundance ($\Zsun$) 5) $\chi^{2}$ / number of degrees of freedom}
\setlength{\extrarowheight}{3pt}
\begin{tabular}[t]{c c c c c}
\hline 
Region & Radius & Temp. & Abund. & $\chi^{2}$/dof \\
\hline
proj$_{1}$ & 3.4--4.2 & $1.3\pm0.2$ & $0.10^{+0.11}_{-0.06}$ & 15/24 \\
proj$_{2}$ & 4.2--4.9 & $2.0^{+0.3}_{-0.2}$ & $0.2^{+0.2}_{-0.1}$ & 26/25 \\
proj$_{3}$ & 4.9--5.7 & $2.7^{+0.4}_{-0.3}$ & $0.06^{+0.16}_{-0.06}$ & 26/25 \\
proj$_{4}$ & 5.7--6.4 & $3.0\pm0.3$ & $0.5^{+0.3}_{-0.2}$ & 26/27 \\
proj$_{5}$ & 6.4--7.1 & $3.5\pm0.3$ & $0.7^{0.4}_{-0.2}$ & 29/26 \\
proj$_{6}$ & 7.1--7.9 & $3.4^{+0.3}_{-0.2}$ & $0.8^{+0.3}_{-0.2}$ & 27/27 \\
proj$_{7}$ & 7.9--9.3 & $3.9\pm0.2$ & $0.5\pm0.1$ & 38/51 \\
proj$_{8}$ & 9.3--10.8 & $4.9^{+0.4}_{-0.3}$ & $0.3\pm0.1$ & 57/50 \\
proj$_{9}$ & 10.8--12.3 & $5.7^{+0.6}_{-0.5}$ & $0.3\pm0.2$ & 22/41 \\
proj$_{10}$ & 12.3--13.8 & $5.7^{+0.6}_{-0.5}$ & $0.8\pm0.2$ & 34/36 \\
proj$_{11}$ & 13.8--15.7 & $7.2^{+0.8}_{-0.7}$ & $0.3\pm0.1$ & 44/44 \\
proj$_{12}$ & 15.7--17.7 & $8\pm1$ & $0.06^{+0.16}_{-0.06}$
& 42/40 \\
proj$_{13}$ & 17.7--19.7 & $7.4^{+0.8}_{-0.7}$ & $0.6\pm0.2$ & 38/39 \\
proj$_{14}$ & 19.7--23.6 & $8.2^{+0.7}_{-0.6}$ & $0.4\pm0.1$ & 90/75 \\
proj$_{15}$ & 23.6--29.5 & $9.0\pm0.6$ & $0.24\pm0.09$ & 119/110 \\
proj$_{16}$ & 29.5--39.4 & $8.3\pm0.4$ & $0.30\pm0.07$ & 166/150 \\
proj$_{17}$ & 39.4--59.0 & $8.6\pm0.4$ & $0.22\pm0.06$ & 206/183 \\
proj$_{18}$ & 59.0--98.4 & $9.7^{+0.6}_{-0.5}$ & $0.28\pm0.08$ & 236/214 \\
proj$_{19}$ & 98.4--147.6 & $12^{+3}_{-2}$ & $0.1^{+0.3}_{-0.1}$ & 195/209 \\
\hline
\end{tabular}
\label{tab:allprojvalues}
\end{table}
\begin{table}
\centering
\caption{Table of best-fitting cluster parameters for each deprojected region.
  Column 1) Region name 2) Inner and outer radii of annulus (\arcsec)
  3) Temperature ($\keV$) 4) Electron density ($\pcmcu$) 5)
  Abundance ($\Zsun$) 6) $\chi^{2}$ / number of degrees of freedom}
\setlength{\extrarowheight}{3pt}
\begin{tabular}[t]{c c c c c c}
\hline 
Region & Radius & Temp. & $n_{e}$ & Abund. & $\chi^{2}$/dof \\
 & & & ($\times10^{-3}$) & & \\
\hline
deproj$_{a}$ & 4.9--8.9 & $2.4\pm0.2$ & $46\pm2$ & $0.4^{+0.2}_{-0.1}$ & 72/58 \\
deproj$_{b}$ & 8.9--14.8 & $4.3^{+0.4}_{-0.3}$ & $26.0^{+0.6}_{-0.5}$ & $0.2\pm0.1$ & 71/58 \\
deproj$_{c}$ & 14.8--24.6 & $7.7^{+1.0}_{-0.8}$ & $11.7\pm0.3$ & $0.5\pm0.2$ & 64/58 \\
deproj$_{d}$ & 24.6--34.4 & $7.7^{+1.0}_{-0.8}$ & $8.8\pm0.2$ & $0.5\pm0.2$ & 66/58 \\
deproj$_{e}$ & 34.4--51.7 & $7.6^{+0.8}_{-0.6}$ & $5.41\pm0.09$ & $0.4\pm0.2$ & 69/58 \\
deproj$_{f}$ & 51.7--64.0 & $10^{+3}_{-1}$ & $3.8\pm0.1$ & $0.2^{+0.3}_{-0.2}$ & 51/58 \\
deproj$_{g}$ & 64.0--88.6 & $8.5^{+1.0}_{-0.8}$ & $2.50\pm0.05$ & $0.3\pm0.2$ & 46/58 \\
deproj$_{h}$ & 88.6--157.4 & $12^{+2}_{-2}$ & $1.08^{+0.02}_{-0.03}$ & $0.1^{+0.3}_{-0.1}$ & 45/58 \\
\hline
\end{tabular}
\label{tab:alldeprojvalues}
\end{table}

\subsection{Deprojected Radial Profiles}
\begin{figure}
\centering
\includegraphics[width=0.95\columnwidth]{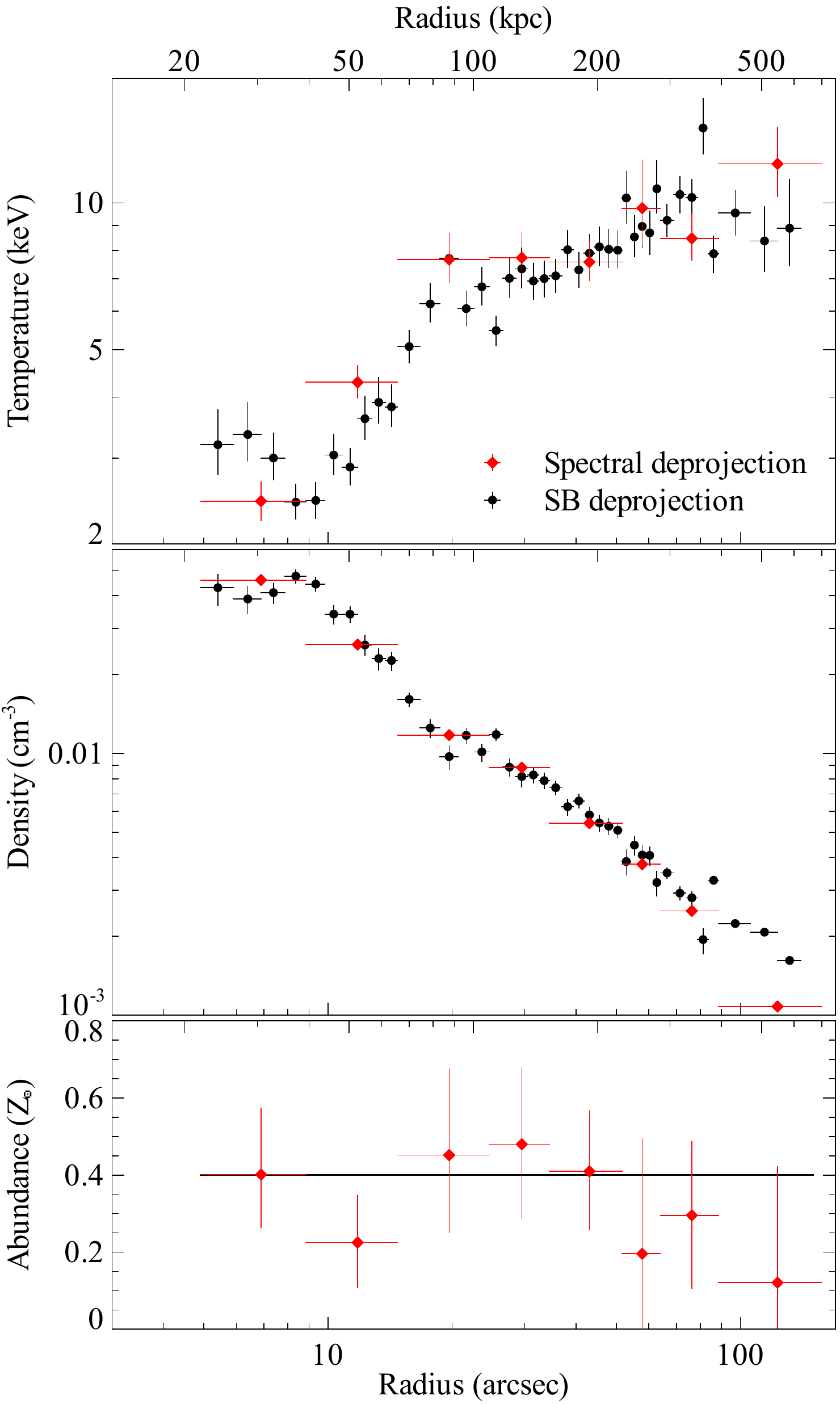}
\caption{Quasar-subtracted deprojected temperature (top), electron density (centre) and
  metallicity (bottom) profiles for the spectral deprojection method
  (red diamonds) and the surface brightness deprojection method (black
  circles).  The radius extends in to 3\arcsec to enable a comparison
  with Fig. \ref{fig:H1821proj}.}
\label{fig:deproj}
\end{figure}

Fig. \ref{fig:deproj} shows the deprojected radial profiles for both
the surface brightness and spectral deprojection methods applied to
H1821+643.  The results for the spectral deprojection are also listed
in Table \ref{tab:alldeprojvalues}.  The surface brightness
deprojection produced a mass profile with a concentration
$c=1.3^{+0.9}_{-0.7}$ and $r_{200}=2.5^{+1.3}_{-0.7}\Mpc$.  The
spatial region from $3-5\arcsec$ was excluded from the deprojection
analysis because the shallow gradient of the core surface brightness
profile and the core cluster substructure produced large errors in the
parameters, in addition to the uncertainty from the quasar subtraction
(Fig. \ref{fig:SBprofile}).

The deprojected temperature profile shows a steep drop inside the
central $100\kpc$, from $10^{+1}_{-3}\keV$ at $\sim250\kpc$ down to
$2.5^{+0.2}_{-0.2}\keV$ in the cluster centre.  The surface brightness
deprojection result is similar but shows an increase in temperature
within the central $40\kpc$.  This appears to correlate with a
decrease in the electron density, calculated from the emission measure
of the deprojected spectrum, however the errors are consistent with a
flat profile.  Although there is a break in the density profile at
$\sim15\arcsec$ corresponding to the surface brightness edge, there
does not appear to be a sharp discontinuity to indicate a shock.  A jump in
either temperature or density cannot be discerned for the
\textsc{dsdeproj} result because large radial bins were required to
ensure enough counts for a good deprojection.  The assumption of
hydrostatic equilibrium in the surface brightness deprojection would
also have smeared out any discontinuity associated with a shock.
Finally, the surface brightness edge is not radially symmetric about
the quasar.  Therefore the circular annuli centred on the quasar that
were used for the deprojection will have inevitably smeared the
spectrum from this region over neighbouring annuli.  Deprojecting the
cluster in the NW sector (Fig. \ref{fig:SBsectors}) where the edge is
strongest and narrowest would be more likely to produce a conclusion
about its nature, however there are not enough X-ray counts to
deproject in sectors.  It was therefore not possible to determine
unequivocally the nature of the surface brightness edge with the
available data.

The metallicity profile produced by spectral deprojection is
consistent with a constant value of $\sim0.4\Zsun$ in the core,
decreasing to $0.2-0.3\Zsun$ at large radii.  The assumption of a
constant abundance of $0.4\Zsun$ for the surface brightness
deprojection was therefore deemed to be suitable.

Both deprojections were repeated with a $\pm10\%$ variation in the
subtracted quasar contribution.  This resulted in an uncertainty of
$\pm0.1\keV$ for the innermost annuli in the surface brightness
deprojection, smaller than the statistical error of $\pm0.3\keV$.  The
variation produced a $\pm0.2\keV$ uncertainty for the innermost bin of
the \textsc{dsdeproj} result but was insignificant at larger radii.
The increase in the central temperature shown in the surface
brightness deprojection is therefore an inconsistency between the two
deprojection methods.  A finer spatial binning over the inner radii
was tried for the spectral deprojection in order to confirm this
feature.  However, the flat surface brightness profile in the core
resulted in a larger fraction of projected emission in each annulus.
Therefore, increasing the number of annuli to be deprojected produced much larger
uncertainties in the central shell.  The increase in core temperature
is not seen in the projected temperature profile (Fig.
\ref{fig:H1821proj}) or the temperature map (Fig.
\ref{fig:jssmaps}).  Whilst the lack of confirmation in projected
profiles or the alternative, low resolution deprojection method does
not preclude an increase in core temperature, the feature could
instead be an artifact of the deprojection method.

The additional assumptions of hydrostatic equilbrium and an NFW mass
model made by the surface brightness deprojection method might not be
appropriate for the cluster core.  For a sample of fourteen galaxy
clusters observed with \emph{Chandra}, \citet{Voigt06} found that the
mass distribution for the majority of the objects was consistent with
the NFW model, however four objects in the sample exhibited a flatter
core.  Forcing an NFW mass model in the cluster core
when the gas mass was in fact much lower could, via hydrostatic
equilibrium, artificially boost the calculated ICM temperature.
Structural features in the cluster core, such as a shock or dense blob
of gas, are also expected to cause spurious deprojection results.

By combining the deprojected temperature $T$ and electron density $n_{e}$, we
calculated derived properties such as the electron pressure
$P_{e}=k_{B}n_{e}T$, radiative gas cooling time 

\begin{equation}
t_{\mathrm{cool}} = \frac{5}{2}\frac{nk_{B}T}{n_en_H\Lambda(T)}
\end{equation}

\noindent where $\Lambda(T)$ is the
cooling function, and entropy $S=k_{B}Tn_{e}^{-2/3}$.  We assumed the temperature and density were
independent of each other in order to calculate errors on the derived
quantities.  The majority of the error was in the temperature values
so this was a reasonable assumption.  

\begin{figure}
\centering
\includegraphics[width=0.9\columnwidth]{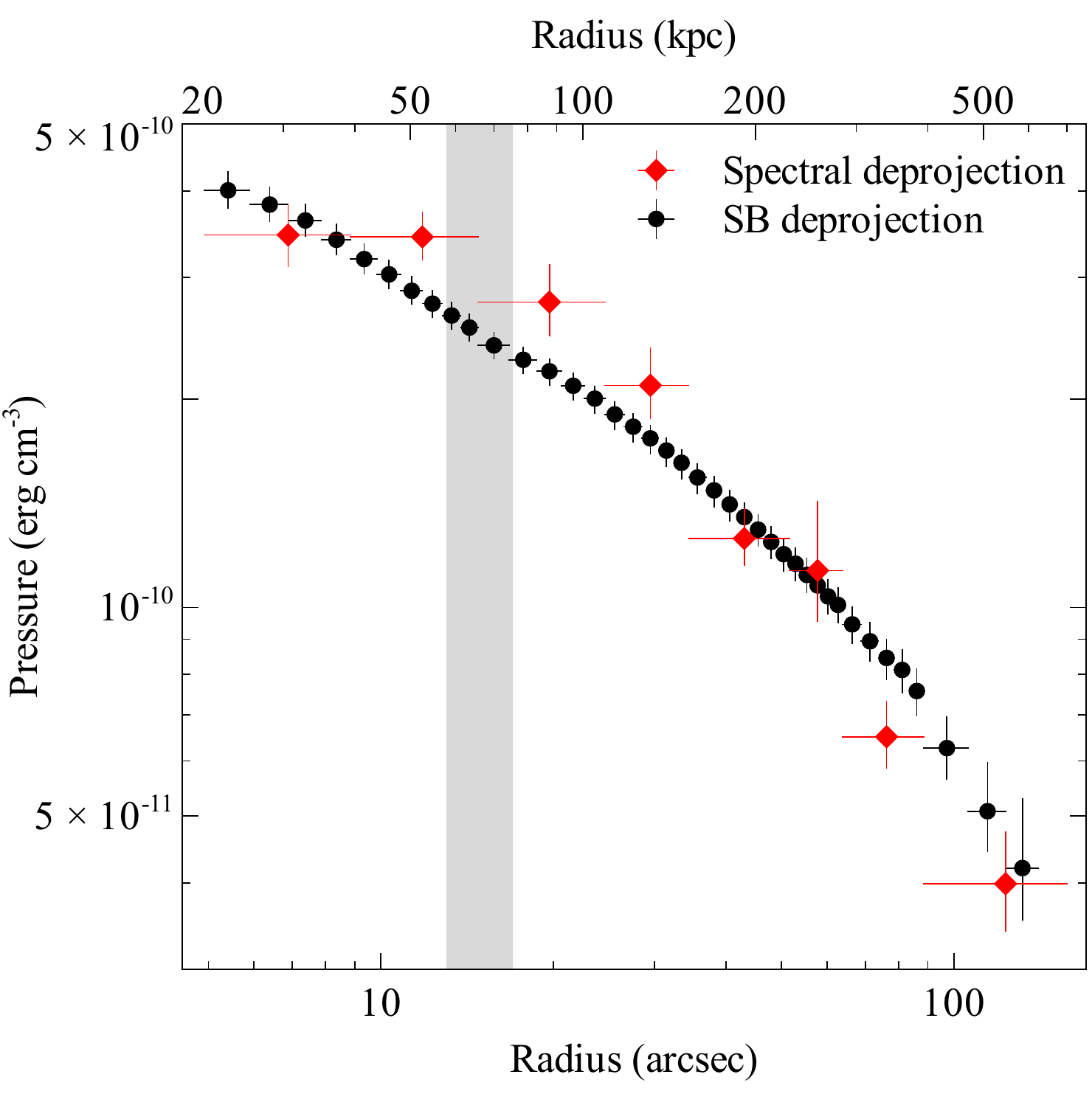}
\caption{Quasar-subtracted deprojected electron pressure profile for the spectral
  deprojection method (red diamonds) and the surface brightness
  deprojection method (black circles).  The shaded region denotes the
  radial extent of the surface brightness edge from $13-17\arcsec$.}
\label{fig:H1821press}
\end{figure}

Fig. \ref{fig:H1821press} shows the electron pressure increases
steadily towards the cluster core.  However, beyond the surface
brightness edge the \textsc{dsdeproj} result flattens possibly to
a constant pressure within $50\kpc$.  The surface brightness
deprojection result assumes hydrostatic equilibrium and an NFW mass
model, which constrain the pressure profile.  Therefore
this conflicting result may indicate that the radio source provides a
significant source of non-thermal pressure in the cluster core.  The
value of the pressure for the innermost radial bin in the spectral
deprojection result should be considered a lower limit on the total
pressure.  It is therefore difficult to determine whether there is a
jump in pressure across the surface brightness edge at $13-17\arcsec$
indicating a shock.  If instead the pressure smoothly changes across
this region the feature could be a cold front (see
\citealt{Markevitch07} and references therein).

\begin{figure}
\centering
\includegraphics[width=0.95\columnwidth]{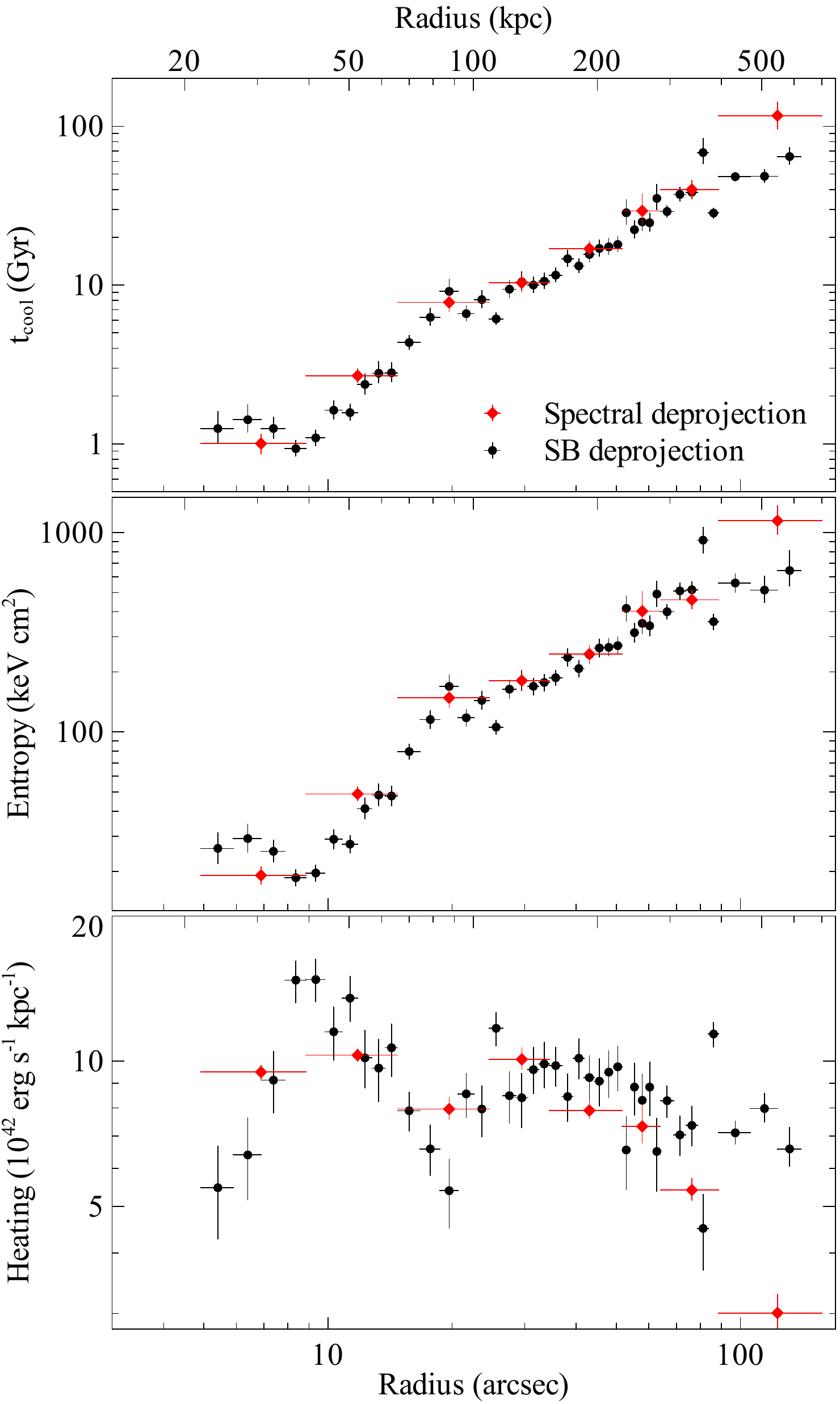}
\caption{Quasar-subtracted deprojected ICM cooling time (top), entropy
  (centre) and heating rate required per $\kpc$ width shell (bottom)
  profiles for the spectral deprojection method (red diamonds) and the
  surface brightness deprojection method (black circles).  The radius
  extends in to 3\arcsec to enable a comparison with
  Fig. \ref{fig:H1821proj}.}
\label{fig:deprojcool}
\end{figure}

The gas cooling time falls steadily down to $1.0^{+0.1}_{-0.1}\Gyr$ at
a radius of 7\arcsec ($30\kpc$) in the core of the cluster
(Fig. \ref{fig:deprojcool}).  With such a short central gas cooling
time, we would expect that gas might be cooling down below X-ray
temperatures.  The cooling radius was defined to be the radius at
which the cooling time fell to $7.7\Gyr$, the time since $z=1$.  The
cooling radius of H1821+643 is at 20\arcsec ($90\kpc$).  The X-ray
mass deposition rate inside the cooling radius was calculated by using
the \textsc{xspec} \textsc{mkcflow} model, which models the emission
from gas cooling between two temperatures where the normalization is
the mass of cooling gas.  

The \textsc{dsdeproj} deprojection process was therefore repeated
using a larger innermost annulus $5-20\arcsec$ ($22-90\kpc$).
However, the majority of the mass deposition in clusters occurs in the
cluster centre, where the radiative gas cooling times are shortest.
The crucial inner $20\kpc$ was necessarily excluded from our analysis,
therefore to get an upper limit on the mass deposition rate, we
assumed a single-phase cooling flow and fitted a
\textsc{phabs(pow)+phabs(mekal+mkcflow)} model to the cooling region
spectrum in \textsc{xspec}.  The lower temperature of the
\textsc{mkcflow} component was therefore fixed to $1\keV$ and the
higher temperature was tied to the temperature parameter in the
\textsc{mekal} component.  The abundance was fixed to $0.4\Zsun$.  The
quasar spectral properties were calculated as before by fitting to the
ChaRT simulation of the cluster spectrum from the cooling region.  The
spectral fit resulted in an upper limit on the mass deposition rate of
$300\pm100\Msunpyr$ within the cooling radius.  Direct detections of
the amount of gas cooling out of the X-ray requires deep high
resolution spectroscopy (eg. \citealt{Peterson03}).

The heating rate refers to the spatial distribution in heat required
to balance the cooling of the cluster gas and was calculated by
dividing the total X-ray luminosity in each shell by the shell width.
Fig. \ref{fig:deprojcool} shows that the rate at which energy must
be deposited in shells of equal width is approximately constant at
$10^{43}\ergpspkpc$ within $100\kpc$, decreases in the radial bin
associated with the surface brightness edge, and then drops off beyond
$200\kpc$.  The total required heating rate to balance the cooling
losses within the cooling radius is therefore $\sim10^{45}\ergps$.  

\subsection{Potential cavity heating}
Assuming the X-ray depression at a distance of $24\kpc$ from the
nucleus is a radio cavity (Fig. \ref{fig:unsharpsub}), we follow the method of
\citet{Dunn04} (see also \citealt{Birzan04}) to calculate the cavity
heating power.  The energy input of the cavity is the sum of the
cavity's internal energy and the work done,

\begin{equation}
E = \frac{1}{\gamma - 1}PV+P\mathrm{d}V \approx \frac{\gamma}{\gamma-1}PV,
\end{equation}

\noindent where $P$ is the thermal pressure of the surrounding ICM, $V$ is the
volume of the cavity and $\gamma$ is the mean adiabatic index of the
fluid in the cavity.  The relativistic case of $\gamma=4/3$ was
assumed.

We used two different methods to place upper and lower limits on the
potential cavity heating power.  For a lower limit, we assumed there
is no additional non-thermal pressure in the cluster core and used the
\textsc{dsdeproj} weak pressure gradient shown in Fig.
\ref{fig:H1821press}.  For an upper limit, the gas pressure profile
from $40-150\kpc$ was extrapolated inwards using a powerlaw to produce an estimate of
the total central pressure, similar to the surface brightness
deprojection result.  The instantaneous mechanical power for the
cavity is $P_{\mathrm{bubble}}=4PV/t_{\mathrm{age}}$, where
$t_{\mathrm{age}}$ is the bubble age.  The cavity was assumed to be
spherical with a radius of $\sim9\kpc$, estimated from the image using
a circular aperture, and located at a distance of
$\sim24\kpc$ from the cluster centre.  Two estimates of the cavity age are
commonly used: the time required for the cavity to rise to its current
location at the speed of sound, $t_{\mathrm{sound}}$, and the time
required for the cavity to rise buoyantly at its terminal velocity,
$t_{\mathrm{buoy}}$.  The sound speed timescale is more applicable to
young radio cavities recently inflated by the AGN.  Its value was
$0.03\Gyr$, which gives a cavity heating power of
$1-2\times10^{44}\ergps$.


Although there could be extra AGN heating power produced by weak
shocks and sound waves (\citealt{FabianPer06}), cavity heating in
H1821+643 is not currently supplying enough heat to replace the
majority of the $\sim10^{45}\ergps$ radiated by the cluster gas.  The
possible `ghost' cavities could potentially produce an additional
power of $9\times10^{44}\ergps$, calculated using the buoyancy timescale.
However, it is not clear from the deprojected surface brightness
profiles (Fig. \ref{fig:deprojSBprofile}) that these features are
cavities.

\section{Discussion}
\subsection{Evidence for quasar and cluster interactions}
\label{sec:qcinteractions}
The morphology in the central $100\kpc$ of H1821+643 suggests a
complex interaction between the quasar and the surrounding cluster
gas.  There are extended arms of emission and an inner radio
cavity with a bright rim in the core and a weak shock or cold front
surrounding most of the cluster core at $\sim15\arcsec$.  However, the
mechanisms by which the quasar output is coupled to the ICM are far
from clear.  

The extended arms of X-ray emission to the N and SE appear to be
correlated with the main axis of the extended FR I radio emission.
These X-ray features may have been pushed or dragged out from the
cluster core by the expanding FR I structure.  Although the inner
radio cavity is too close to the quasar to be analysed effectively,
this indicates that there could be some mechanical feedback in the
cluster core.  The ends of the FR I structure appear to terminate at
the surface brightness edge seen in the X-ray so this feature could be a
weak shock generated by the radio source expanding into surrounding
cluster gas.  However, \citet{Blundell01} found that the outermost ends of the
extended radio emission do not appear to end in shocks.  They
inferred that the extended emission has not grown any faster than the
ambient sound speed of the ICM.  In addition, the extended X-ray
structure to the W is not associated with any radio emission and the
surface brightness edge appears strongest in this radial
direction (Fig. \ref{fig:deprojSBprofile}).  

Based on optical imaging, \citet{HutchingsNeff91} suggested that
H1821+643 is in the late stages of a mild tidal event or merger,
although there is no sign of a disturbing object.  \citet{Fried98}
found that the line widths of the extended emission-line gas in
H1821+643 produced velocity dispersions in agreement with this
interaction scenario.  However, velocity dispersions of a few hundred
$\kmps$ are also consistent with the kinematics of emission-line
nebulae in cool-core clusters (\citealt{Hatch07}).  The X-ray features
could be interpreted as resulting from a merger interaction which has
disturbed the gas in the cluster core.  \citet{Blundell01} estimated
that if the radio plumes have emerged no faster than the ambient sound
speed then the radio structure has been stable for a minimum of
$3\times10^{8}\yr$.  The surface brightness edge could be a weak shock
generated by a merger at that time or more likely, given that it can
be observed around most of the cluster core, it could be a cold front
produced by the cool-core gas sloshing in the cluster potential well
(\citealt{Markevitch00}; \citealt{Vikhlinin01}).  The swirl feature
seen in the projected temperature map would also support this
interpretation (Fig. \ref{fig:jssmaps}).

H1821+643 shows no obvious evidence for X-ray absorbing winds from the
quasar in the \emph{Chandra} HETG (\citealt{Fang02}), LETG
(\citealt{Mathur03}) or in \emph{FUSE} spectra (\citealt{Oegerle00}).  However,
outflows could be oriented in the plane of the sky and therefore not
evident along our line of sight.  \citet{Oegerle00} observed
associated absorption at the redshift of H1821+643 in the \emph{FUSE}
and \emph{HST} spectra (see also \citealt{Bahcall92}) and concluded
that this was unlikely to be quasar intrinsic absorption.  Instead
they proposed that the broad range of ionization present required a
multiphase absorber that is likely located in the central cluster
galaxy.  We found that the X-ray spectra for the central two annuli,
that cover the spatial extent of the central galaxy, did not require
any additional absorption above the Galactic column density determined
by \citet{Kalberla05}.  However, these annuli contain a significant
fraction of the quasar PSF and any additional absorption contribution
to the spectrum was difficult to determine.  We compared soft
($0.5-1.0\keV$ and $1.0-2.0\keV$) and hard band ($3.0-7.0\keV$) images
to see if there were any obvious potential sites for absorption.  The
X-ray depression at $24\kpc$ from the nucleus was clearly visible in
both soft band images, surrounded by cool rims of gas, however it was
not visible in the hard band image.  The hard band image contains far fewer counts
and was dominated by the quasar PSF.  Therefore a small cavity would
be particularly difficult to distinguish.  However, this could also be
interpreted as a cool extended arm of emission, similar to those in
the N and NW directions, with a superimposed absorption region.

\subsection{Accretion Mechanisms}
\label{sec:accretion}
The total energy output, radiative and mechanical, of the AGN can be
used to infer an accretion rate for H1821+643, assuming an accretion
efficiency for each mechanism.  Using a bolometric correction of 50
from \citet{Vasudevan07} with the quasar luminosity in the energy
range $2-10\keV$ $L_X=(4.2\pm0.1)\times10^{45}\ergps$, we calculated the
quasar bolometric luminosity $L_{\mathrm{bol}}\sim2\times10^{47}\ergps$.
However, this could have been underestimated if a significant number
of X-ray photons from the quasar were scattered by the ICM.  By
extrapolating the electron density profile in to $\sim0.1\pcmcu$, we
estimated an optical depth for the central $30\kpc$ region of
$\tau\sim0.002$.  Therefore, only $\sim1$ in 1000 photons from the
quasar would be scattered by the ICM and this effect is
insignificant.  

For a radiative efficiency $\epsilon=0.1$, the mass accretion rate
required to power the quasar is therefore
$\dot{M}_{\mathrm{acc}}=L_{\mathrm{bol}}/\epsilon c^2 \sim40\Msunpyr$.  The mechanical
power calculated by cavity heating is several orders of magnitude
lower than the quasar luminosity and therefore is not included in an
estimate of the required accretion rate.  The calculated mass
accretion rate for H1821+643 can be compared with the theoretical
Eddington and Bondi accretion rates.

The Eddington accretion rate can be expressed as 

\begin{equation}
\frac{\dot{M}_{\mathrm{Edd}}}{\Msunpyr}=\frac{6.6}{\epsilon}\left(\frac{M_{BH}}{3\times10^9\Msun}\right)
\end{equation}

Using the H$\alpha$, H$\beta$ and MgII line widths and the optical/UV
continuum luminosity (\citealt{Kolman93}), together with the recipes
in \citet{McGill08} (their section 4.1 and references therein) we estimated a black hole
mass $M_{BH}\sim3\times10^{9}\Msun$ for H1821+643.  The uncertainties
in the bolometric correction and the black hole mass were at least
50\% and therefore subsequent calculations were assumed to be only an
estimate.  Using the estimate of black hole mass, the Eddington
luminosity was calculated to be
$L_{\mathrm{Edd}}\sim4\times10^{47}\ergps$, giving an Eddington accretion rate
$\dot{M}_{\mathrm{Edd}}\sim70\Msunpyr$ and a high Eddington ratio
$L_{\mathrm{bol}}/L_{\mathrm{Edd}}\sim0.5$.

Assuming spherical symmetry and negligible angular momentum, the Bondi
rate is the rate of accretion for a black hole with an accreting
atmosphere of temperature, $T$, and density, $n_e$,
(\citealt{Bondi52}) and can be expressed as

\begin{equation}
\frac{\dot{M}_{\mathrm{Bondi}}}{\Msunpyr}=0.12\left(\frac{k_{B}T}{\keV}\right)^{-3/2}\left(\frac{n_e}{\pcmcu}\right)\left(\frac{M_{BH}}{3\times10^9\Msun}\right)^2
\end{equation}

\noindent for an adiabatic index $\gamma=5/3$.  This accretion occurs
within the Bondi radius, $r_{\mathrm{Bondi}}$, within which the gravitational
potential of the black hole dominates over the thermal energy of the
surrounding gas,

\begin{equation}
\frac{r_{\mathrm{Bondi}}}{\kpc}=0.10\left(\frac{k_BT}{\keV}\right)^{-1}\left(\frac{M_{BH}}{3\times10^9\Msun}\right).
\end{equation}

The Bondi accretion rate is therefore an estimate of the rate of
accretion from the hot ICM directly onto the black hole.  The Bondi
radius is not resolved for H1821+643 so the electron density and
temperature from the innermost annulus were used as lower and upper
limits, respectively, on the accretion atmosphere properties producing
an underestimate of the Bondi accretion rate.  For an
innermost temperature $2.4\pm0.2\keV$ and electron density
$0.046\pm0.002\pcmcu$, the Bondi accretion rate is estimated to be
$\dot{M}_{\mathrm{Bondi}}\sim0.001\Msunpyr$ at the Bondi radius
$r_{\mathrm{Bondi}}\sim40\pc$.  Although this estimate of the Bondi accretion
rate is expected to lie below the true value, the implied accretion
rate for the quasar exceeds this by a factor of $\sim10^4$.  

Following \citet{Allen06}, we also estimated the Bondi accretion rate
by extrapolating the temperature and density of the cluster gas in to
the Bondi radius.  This was achieved by using powerlaw fits to the
inner annuli of the \textsc{dsdeproj} deprojected temperature and density profiles
in Fig. \ref{fig:deproj}.  The steep
temperature ($T(r)\propto r^{0.6}$) and density gradients
($n_e(r)\propto r^{-1.2}$) in the cool-core produced an upper limit on
the Bondi accretion rate of $\dot{M}_{\mathrm{Bondi}}\sim6\Msunpyr$ at
$r_{\mathrm{Bondi}}\sim0.5\kpc$.  However, an extrapolation over two orders of
magnitude in radius is unlikely to provide a good estimate of the
Bondi accretion rate.  \citet{Allen06} analysed elliptical
galaxies, which had shallow central temperature gradients and X-ray
data permitting a measurement of the gas properties within one order
of magnitude of the Bondi radius.  

Compton cooling of the accretion flow by radiation from the quasar
could potentially increase the flow of material from a hot atmosphere
onto the black hole (\citealt{FabianCrawford90}).  Since Compton
scattering between strong UV photons and hot electrons can dominate
the cooling process for the ICM in the cluster core, accretion by
Compton cooling can supply gas to the central black hole and, as shown
by \citet{FabianCrawford90}, the accretion rate by Compton cooling
flow can be larger than that of Bondi accretion alone.  Using the
cluster gas properties in the innermost annulus, we determined that
Compton cooling will dominate bremsstrahlung cooling within a radius
$R_C\sim5\kpc$ and rapidly cool the gas down to the Compton
temperature.  

The Compton temperature was calculated using the quasar spectral
energy distribution (SED) from the optical to the soft X-ray presented
in \citet{Kolman93}.  We fitted the SED in \textsc{xspec} with their
Kerr disk model for the UV bump (Fig. 16(c) in \citealt{Kolman93})
and a broken power-law to provide an extension up to $100\keV$, as
described in \citet{Vasudevan09}.  Therefore, with the bulk of the
emitted quasar luminosity in the UV, the Compton temperature is
$T_{C}\sim0.4\keV$.  

If the accreting material is quickly Compton cooled within
$\sim5\kpc$, as it approaches the Bondi radius, then the Bondi radius
will grow from $r_{\mathrm{Bondi}}\sim40\pc$ to $\sim250\pc$ and the
Bondi accretion rate increases from
$\dot{M}_{\mathrm{Bondi}}\sim0.001\Msunpyr$ to $\sim6\Msunpyr$.
Therefore, Bondi accretion of rapidly Compton cooled material could
provide a significant source of fuel for the quasar.

Cold accretion mechanisms, such as the cold feedback scenario
suggested by \citet{PizzolatoSoker05}, could provide an alternative
route for accreting material.  Within the cooling region of H1821+643,
up to $300\pm100\Msunpyr$ could be cooling down below X-ray temperatures.
\citet{PizzolatoSoker05} proposed that overdense blobs of cool gas
from within the entire cooling region rapidly cool and sink towards
the central black hole or condense to form cold molecular clouds and
stars.  A large influx of material onto the black hole, caused by an
accretion disk instability for example, would then trigger an
outburst.  For cold accretion to provide a source of fuel for the
quasar, a significant fraction of the gas cooling out of the X-ray
must be funnelled down onto the black hole.  Such an efficient, steady
accretion of material from within the cooling region seems unlikely,
however the cooling material could have built up in the cluster centre
to form a large reservoir of cool gas which feeds the nucleus after
a triggering event.

Alternatively, a merger scenario could have initiated the quasar
outburst by providing a rapid influx of fuel for the central engine.
The precession in the jet axes suggested by the observations of
\citet{Blundell01} could have arisen if the central engine was
composed of two orbiting black holes that had not yet merged.  A
merger scenario could therefore potentially have triggered the quasar
outburst and the jet precession.  However, the galaxy cluster's deep
gravitational potential will tidally strip a merging subcluster of its
gas in the cluster outskirts, making it difficult to channel the
merging material effectively down onto the central black hole.

\subsection{Impact of the quasar on the galaxy cluster core}
\label{sec:quasarimpact}
\begin{figure}
\centering
\includegraphics[width=0.9\columnwidth]{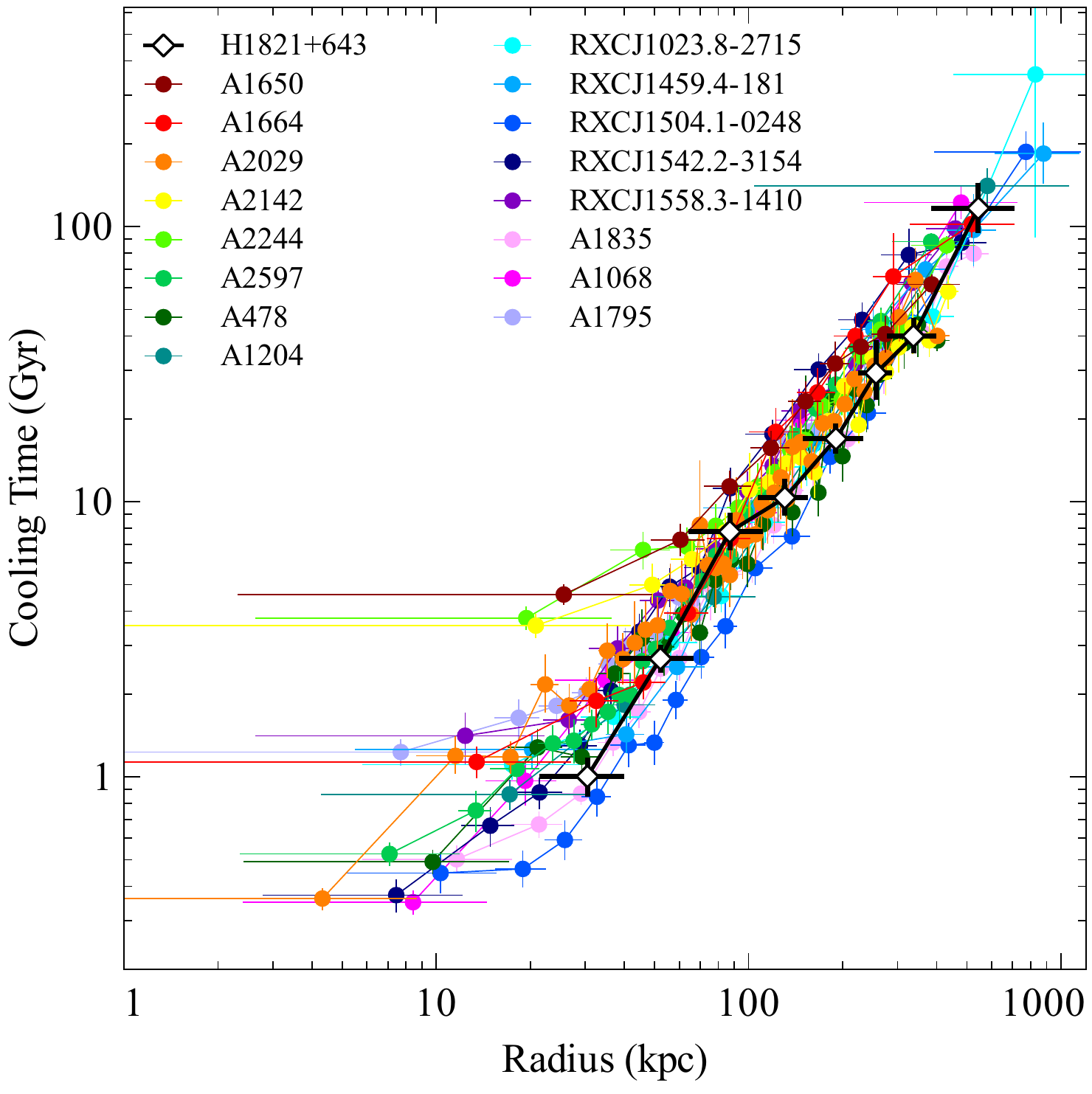}
\caption{\textsc{dsdeproj} deprojected gas cooling times for a sample
  of cool-core and non-cool-core galaxy clusters.}
\label{fig:tcool_all}
\end{figure}

Fig. \ref{fig:tcool_all} compares the radiative gas cooling time
profiles of H1821+643 with a sample of 16 cool-core and non-cool-core
clusters with \emph{Chandra} archival observations.  The clusters were
roughly selected on redshift ($z=0.06-0.25$) and the number of X-ray counts in the
\emph{Chandra} observation.  The first criterion ensured a similar
intrinsic spatial extent on the ACIS chip to H1821+643 and the second
provided an equivalent resolution with at least eight deprojected
radial bins ($\sim40,000$ counts).  In addition, three non-cool-core clusters, Abell 1650,
Abell 2142 and Abell 2244, were included to serve as a contrast.

H1821+643 has a cooling time profile typical of a strong cool-core
cluster with a more passive central engine, such as Abell 1835.
Underneath the quasar, the large-scale properties of the
galaxy cluster have not been strongly affected and the ICM is continuing to
evolve in a similar way to other cool-core clusters.

This period of quasar activity in H1821+643 is unlikely to be
long-lived.  For such a high accretion rate of $40\Msunpyr$, the
e-folding time of the SMBH growth, assuming a radiative efficiency
$\epsilon=0.1$, is $\sim10^8\yr$ and in this time the SMBH would need
to accrete $\sim4\times10^9\Msun$ of material.  A period of rapid
black hole growth longer than $\sim10^8\yr$ would therefore result in
a black hole mass far larger than the typical upper limit observed
(eg. \citealt{Marconi04} and references therein).  On this timescale,
we would not expect radiative heating from the quasar to
significantly alter the gas cooling time calculated in large radial
bins.  However, to quantify the effect of the active nucleus on the
cluster gas we have run Cloudy (version 08.00) (\citealt{Ferland98})
models appropriate to conditions in the regions $15-30\kpc$
(projected) and $22-39\kpc$ (\textsc{dsdeproj} deprojected) from the
nucleus.

The Cloudy simulations are described in detail in Appendix B.  In
summary, we generated the intrinsic continuum shape for the active
nucleus using the quasar SED model described in section
\ref{sec:accretion}.  A model of the cluster gas, with suitable temperature
$2.4\keV$ and metallicity $0.4\Zsun$, at the distance of the extracted
X-ray spectrum was then irradiated with this quasar continuum.  The
emitted spectrum of the Cloudy model, together with a model accounting
for the quasar PSF, were fitted to the observed spectra from the
projected ($15-30\kpc$) and deprojected ($22-39\kpc$) regions.  We
also produced a Cloudy model for only the cluster emission to provide
a simple comparison for the Cloudy model with the added quasar
contribution.

\begin{figure}
\centering
\includegraphics[width=0.9\columnwidth]{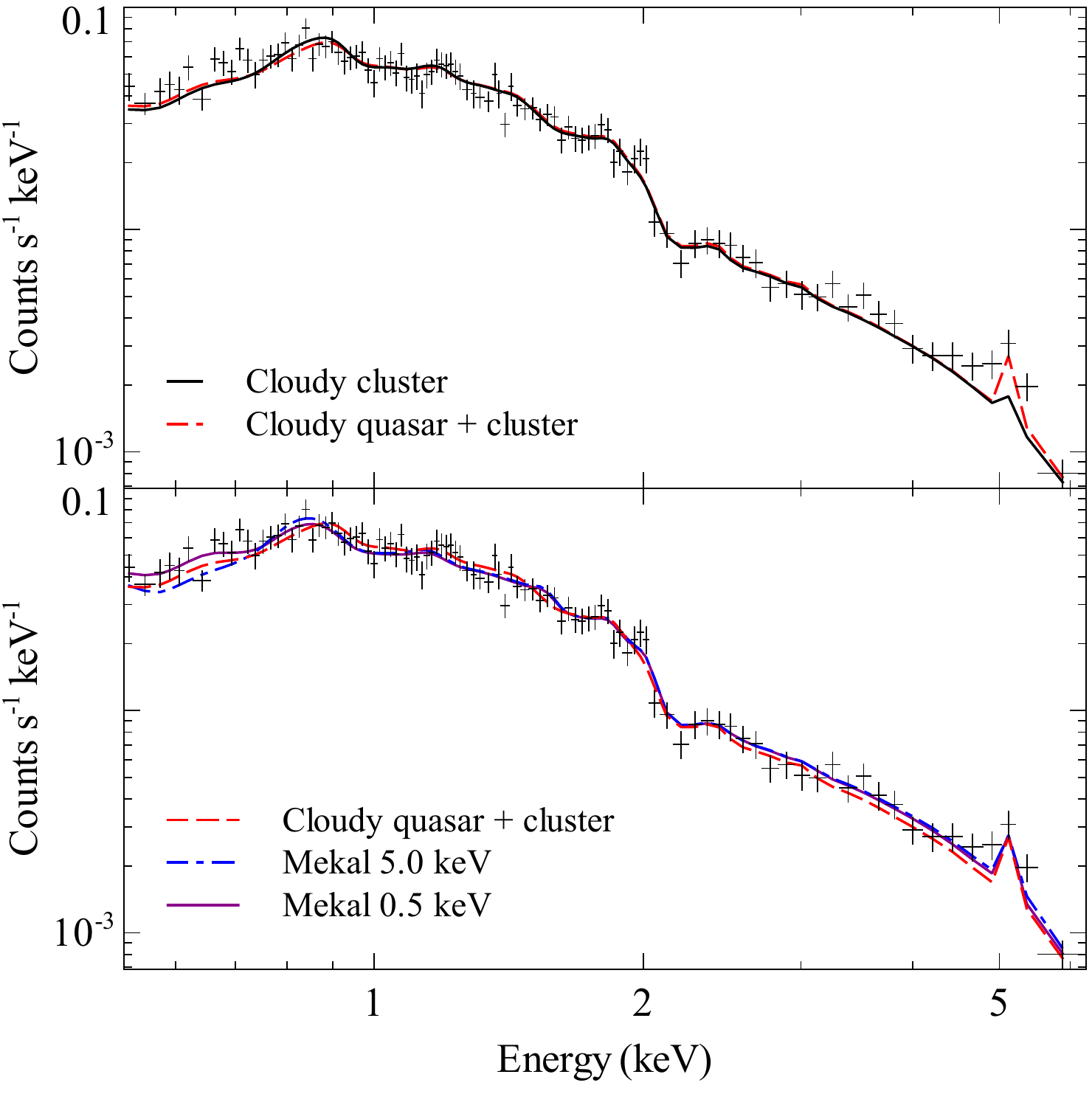}
\caption{The cluster projected spectrum from $15-30\kpc$ is shown with
  the best-fit Cloudy and \textsc{xspec} models superimposed.  Upper:
  the Cloudy models for the cluster gas (black solid line) and the
  cluster gas irradiated by the quasar (red dashed line).  Lower: the
  Cloudy model irradiated by the quasar (red dashed line) and the
  two-temperature \textsc{mekal} model with a $5.0\keV$ (blue
  dot-dashed line) or a $0.5\keV$ (purple solid line) second
  temperature component.  The contribution of the quasar PSF was
  included as a \textsc{phabs(powerlaw)} component in each model.  The
  three models produce a very similar spectral fit.}
\label{fig:cloudymodels}
\end{figure}

The addition of the Cloudy quasar component did not significantly
alter the shape of the model spectrum (Fig. \ref{fig:cloudymodels}).
However in this model, photoionization of the cluster gas produced a
significantly improved fit to
the Fe\,K\,$\alpha$ line blend at $5\keV$ in the projected spectrum
($\chisq=130$ reduced to $\chisq=110$ for 100 degrees of freedom).
This improvement might be due to continuum pumping (\citealt{Porter07}).

This could provide an explanation for the sudden drop in metallicity
seen between 15 and $26\kpc$ in the projected profile
(Fig. \ref{fig:H1821proj}).  This effect will obviously be stronger
closer to the nucleus, therefore it is unlikely to be observed in
the deprojected metallicity profile which did not extend as close in
to the quasar as the projected profile.

However, whilst our Cloudy simulations suggest that the quasar could
be significantly photoionizing the gas close to the nucleus, there are
several other possible explanations for the observed drop in core
metallicity.  \citet{Buote94} argue that fitting intrinsically
multitemperature spectra with single-temperature models artificially
produces a lower metallicity value.  There could be a significant
contribution from a second temperature component in the observed
projected spectrum of H1821+643, such as a hotter projected cluster
emission or emission from cooling gas in the cluster core.  

We therefore added a second temperature component to our
\textsc{xspec} model to give a \textsc{phabs(powerlaw)+phabs(mekal+mekal)}
model which was fitted to the projected spectrum ($15-30\kpc$).  The
second temperature component was first fixed to $5.0\keV$ and the
metallicity in both temperature components was fixed to $0.5\Zsun$.
Fig. \ref{fig:cloudymodels} (lower panel) shows that the addition of a second
temperature component produced an improved fit to the Fe\,K\,$\alpha$
line blend.  However, a cooler second temperature component of
$0.5\keV$ was found to provide a significantly better fit to the
projected spectrum than the $5.0\keV$ component ($\chisq=94$ for
$0.5\keV$ compared to $\chisq=109$ for $5.0\keV$ with 99 degrees of
freedom).  Fig. \ref{fig:cloudymodels} suggests that a second
temperature component, hotter or colder, or photoionization of the
cluster gas could explain the poor fit to the Fe\,K\,$\alpha$
line blend.

Assuming pressure equilibrium, the volume filling fraction of the
$0.5\keV$ temperature component $i$ is

\begin{equation}
f_{i} =
\frac{\varepsilon_{i}T_{i}^2}{\sum_{j}\varepsilon_{j}T_{j}^{2}}
\end{equation}

\noindent where $\varepsilon_{i}$ is the emission measure
(\textsc{xspec} normalization) of the component $i$, $T_{i}$ is the
temperature and $j$ sums over all the components.  The $0.5\keV$
temperature component has a volume filling fraction of $0.2\%$, which
equates to a gas mass of $M_{gas}\sim3\times10^9\Msun$.  With a very
short radiative gas cooling time of only $50\Myr$, the $0.5\keV$
temperature component could produce a mass deposition rate of
$\sim50\Msunpyr$ in the central $30\kpc$ in the absence of heating.

Finally, an additional power-law component from inverse-Compton
emission could also provide an explanation for the poor fit to
Fe\,K\,$\alpha$.  A power-law component would raise the level of the
continuum emission in the projected spectrum and reduce the prominence
of the emission lines, in a similar way to the extra $5\keV$
temperature component.  However, coincident cluster emission and
quasar PSF made it difficult to put any constraints on inverse-Compton
emission and therefore we cannot determine whether this would resolve the
model fit.

Although from our results it is not possible to determine a preferred
explanation for the poor fit to the Fe\,K\,$\alpha$ line blend, we
argue that a combination of the above would be most likely.  The
projected spectrum will contain a contribution of hotter
($5.0-8.0\keV$) gas from the outer cluster layers.  Our Cloudy
simulations suggest that the quasar emission should produce a
significant amount of photoionization in the central $30\kpc$.
Observations of strong cool-core clusters with the \emph{XMM-Newton}
Reflection Grating Spectrometer have confirmed that many have a second
cooler temperature component in the central region
(\citealt{Peterson03}; \citealt{Kaastra04}; \citealt{Sanders08}).
Finally, although we are unable to quantify any inverse-Compton
emission, this could also be a factor close in to the nucleus. 

\subsection{AGN Unified Accretion Model}
\citet{Churazov05} (see also
\citealt{DiMatteo05}; \citealt{Croton06}) argue that a parallel can be drawn
between the evolution of Galactic black holes (eg. \citealt{Fender99};
\citealt{Gallo03}) and SMBHs (\citealt{Merloni03};
\citealt{Maccarone03}; \citealt{Falcke04}; \citealt{Chiaberge05}).
SMBHs and their surrounding medium are therefore expected to evolve through two
stages.  Early on, the SMBH grows rapidly by accreting cooling gas
near the Eddington rate with weak feedback on the surrounding medium
but a very high radiation efficiency (eg. \citealt{YuTremaine02}).
This is known as the `quasar-mode' which terminates when, despite the
weak coupling to the gas, the SMBH produces enough heat to suppress
cooling.  With a lower accretion rate, the system moves to a more
passively evolving stage, `radio-mode', where efficient mechanical
feedback sustains the ICM at X-ray temperatures.  Recent observations
of nearby low-luminosity SMBHs in clusters and galaxies suggest that
mechanical feedback from the AGN, in the form of bubbles, shocks and
sound waves, is effective at heating the cluster gas
(eg. \citealt{FabianPer03}; \citealt{Birzan04};
\citealt{DunnFabian06}; \citealt{Rafferty06}; \citealt{FabianPer06}).

With rapid accretion at half of the Eddington rate, H1821+643 is
currently in `quasar-mode', corresponding to the high/soft state of
black hole X-ray binaries where a geometrically thin, optically thick
accretion disk forms (\citealt{Shakura73}).  However, quantifying the
quasar feedback into the surrounding ICM in H1821+643 is difficult.
H1821+643 has a gas cooling time profile typical of a strong cool-core
cluster with a passive central engine indicating that the quasar has
not strongly affected the large-scale cluster gas properties.  The
Cloudy simulations detailed in section \ref{sec:quasarimpact} also
suggest that the radiative output of the luminous quasar has little
impact on the bulk of the cluster gas.  There is some indication of
mechanical heating, although this has a low heating power and would
only currently compensate for $\sim10\%$ of the ICM cooling losses.
As discussed in section \ref{sec:qcinteractions}, there is no obvious
evidence for X-ray absorbing winds in H1821+643 in the existing
observations.  However, outflows might not be evident along our line
of sight and therefore cannot be ruled out.  Although the cluster gas
properties do not indicate any efficient feedback from quasar winds on
the scales considered ($r>15\kpc$) this does not exclude an efficient coupling to
the cold galaxy gas (see \citealt{Chartas02,Chartas03,Chartas07};
\citealt{Pounds03}; \citealt{Reeves09}).

H1821+643 could provide an example of the early stage of
evolution for these systems where the SMBH has not evolved beyond
`quasar-mode' and cannot produce enough heat to suppress cooling in
this rich cluster.  The underlying cluster may exhibit the early
cool-core environment before `radio-mode' heating switches on to
efficiently reheat the gas.  It seems implausible that the quasar
could remain switched on for longer than $\sim10^8\yr$ because in that
time the black hole mass will increase to $\sim10^{10}\Msun$, greater
than the typical observed upper limit (eg. \citealt{Marconi04}).

However, the FR I radio structure shown by H1821+643 appears to be at
odds with the simple picture of `quasar-mode' feedback.  Around 10\%
of quasars are known to have strong radio emission
(eg. \citealt{Ivezic02}) indicating that SMBHs with high accretion
rates can be associated with powerful outflows.  By drawing
comparisons with observations of microquasars, \citet{Nipoti05}
suggest a parallel between radio-loudness in quasars and the flaring
mode in microquasars, where radio-loudness is simply defined as
extended radio emission (FR I or FR II).  Microquasars have been
observed to have two distinct modes of energy output: one producing
roughly steady radio emission that is coupled with X-ray emission
(coupled mode), the other producing strongly variable, flaring radio
emission that appears to be decoupled from the X-ray emission (flaring
mode).  \citet{Nipoti05} further associate radio-quiet quasars with
non-flaring/coupled states of microquasars, in either low/hard or
high/soft states.  

In this picture, H1821+643 would be classed as a radio-loud quasar in
flaring mode, where the nucleus is producing the extended FR I radio
structure.  These flares could be analogous to the short, radio
outbursts that have been observed in transient X-ray binaries during
the transition from the low/hard to high/soft state when the accretion
rate approaches the Eddington limit (\citealt{Fender04}).  However,
whilst black hole accretion can be scaled from microquasars and X-ray
binaries up to quasars, the environments of these objects vary
significantly.  It is unclear how the rich cluster environments
and accretion history of quasars will affect the conclusions
drawn from smaller scale systems. 

\section{Conclusions}
By accurately subtracting the quasar contribution, we were able to
determine the properties of the surrounding ICM down to $3\arcsec$
from the quasar.  The temperature of the cluster gas decreases from
$9.0\pm0.5\keV$ beyond $\sim200\kpc$ down to $1.3\pm0.2\keV$ inside
$20\kpc$, with a short central radiative cooling time of
$1.0\pm0.1\Gyr$, typical of a strong cool-core cluster.  By comparing
the cooling time with a sample of cool-core clusters with `radio-mode'
AGN, we determined that the quasar does not appear to have had a
significant impact on the large-scale cluster gas properties.

The cluster core features several extended arms of emission, which
could have been dragged out by the expanding FR I structure.  However,
we found that the radio emission is not clearly related to the X-ray
gas morphology.  The nature of the surface brightness edge at
$\sim15\arcsec$ could not be unequivocably determined from the
temperature and density profiles.  This could be interpreted as a weak shock
generated by the expansion of the radio lobes, or a cold front
produced by the cool-core sloshing in the cluster's
gravitational potential.  There is some evidence for cavity heating in
the cluster core which could currently compensate for $\sim10\%$ of
the ICM cooling losses.

The derived accretion rate of $40\Msunpyr$ is approximately half of
the quasar's Eddington limit.  We showed that Bondi accretion, boosted
by Compton cooling of the accretion material by the quasar radiation,
could provide a significant proportion of the required fuel.  However,
without resolving the cluster gas properties at the Bondi radius, we
could not determine whether this mechanism could supply all of the
required $40\Msunpyr$ and potentially provide a self-sustaining fuel
source for the quasar.  We have also discussed the alternative
mechanisms of cold accretion or a rapid infall of a large quantity
cool gas provided by a subcluster merger.  

In the context of AGN unified accretion models, we suggest that
H1821+643 could provide an example of the early stage of evolution for
these systems where the SMBH has not evolved beyond `quasar-mode' and
cannot produce enough heat to suppress cooling.  The surrounding rich
cluster could therefore exhibit the early cool-core environment before
the onset of `radio-mode' heating efficiently reheats the gas.
However, the powerful FR I radio structure observed in H1821+643, and
in around 10\% of quasars (eg. \citealt{Ivezic02}), conflicts with
this simple scenario of `quasar-mode' feedback.  Following the
comparisons with observations of microquasars suggested by
\citet{Nipoti05}, we consider that H1821+643 is in a flaring mode
where the radio emission is decoupled from the X-ray emission.  These
flares could be analogous to the short, radio bursts observed in
transient X-ray binaries during the transition from the low/hard to
high/soft state (\citealt{Fender04}).  However, it is unclear how the
conclusions drawn from the smaller scale systems will be affected by
the rich cluster environment and the quasar accretion history.

\section*{Acknowledgements}
We acknowledge support from the Science and Technology Facilities
Council (HRR), the Royal Society (ACF, KMB) and from CXC Grant
GO8--9121X (WNB).  We thank Gary Ferland for helpful discussions.  We
also thank Mark Bautz for suggesting the use of the ACIS readout
streak to extract a source spectrum and the \emph{Chandra} X-ray
Center for the analysis and extensive documentation available on the
\emph{Chandra} PSF.  We thank the referee for helpful comments.

\section*{Appendix A: Simulating the Quasar PSF}
\subsection*{A1 Readout streak spectrum}
For heavily piled up sources, M. Bautz suggested using the ACIS
readout streak to extract a source spectrum (\citealt{Gaetz04}).  In this
observation, ACIS accumulates events for a frame exposure time of
$3.1\s$ and then reads out the frame at a parallel transfer rate of
$40\mus$, taking a total of $0.04104\s$ to read out the entire frame.
During readout, the CCD still accumulates events, which are
distributed along the whole column as it is read out, creating a
continuous streak for very bright point sources.

\begin{figure}
\centering
\includegraphics[width=\columnwidth]{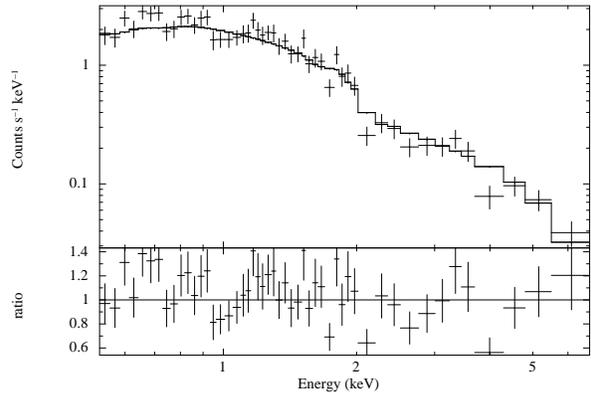}
\caption{H1821+643 readout streak spectrum and spectral fit.  The spectrum has
  been scaled by the ratio of the observation exposure time to the
  effective streak exposure time.}
\label{fig:readoutspectrum}
\end{figure}

We extracted a spectrum in the energy range $0.5-7.0\keV$ from two
narrow regions ($300 \times 8\pix^2$) of the readout streak positioned
either side of the central source and avoiding as much of the cluster
emission as possible.  Any remaining cluster emission in the spectrum
was then subtracted by using background regions adjacent to the
readout streak.  Using a wider extraction region for the readout
streak was found to increase the uncertainty in the spectrum without
altering the results.


The accumulated exposure time in the streak during frame transfer was
$ (300 \times 2) \times 27463 \times 40\mus = 659.11s$, where 27463
was the total number of frames.  Therefore, the readout spectrum count
rate was multiplied by a factor of 129.17, the ratio of the
observation exposure time to the readout streak exposure time.
Previous analyses of \emph{Chandra} readout streak data have indicated
that a gain correction of $+2.5\%$ for the ACIS-S3 chip
(\citealt{Gaetz04}).  Frame transfer is effectively a different ACIS mode for
which there is likely to be a different response to arriving events
compared to the calibrate timed exposure mode.  However, for the
H1821+643 readout streak on the ACIS-S3 chip, we found that the
recommended $+2.5\%$ gain correction did not significantly alter the
spectral fitting results and have not applied it to our analysis.

The response files for the transfer streak spectrum were generated
using the standard \textsc{ciao} tools \textsc{mkacisrmf} and
\textsc{mkwarf}.  The response matrix (RMF) was extracted over the
readout streak region used in this analysis.  However, the effective
area (ARF) was determined using a region at the location of the quasar
image.  The X-rays in the transfer streak actually hit the detector at
the direct image, and their detected position is an artifact of the
chip readout.  The scaled readout streak spectrum was grouped with a minimum
of 50 counts per spectral bin, corrected for the applied count rate
scaling factor.

The scaled readout streak spectrum, shown in Fig.
\ref{fig:readoutspectrum}, was fitted with an absorbed power-law model
in the spectral fitting package \textsc{xspec} version 12.  The best
fit photon index was consistent with the \citet{Fang02} analysis of
the \emph{Chandra} HEG and MEG spectra and the previous \emph{ASCA}
result (\citealt{Yamashita97}).  Therefore, we fixed the photon index
to the more accurate HEG/MEG result, $\Gamma=1.761^{+0.047}_{-0.052}$
(\citealt{Fang02}).  The Galactic absorption parameter was left free,
giving a value of $0.003\times10^{22}\pcmsq$ with an upper limit of
$0.04\times10^{22}\pcmsq$.  \citet{Fang02} inferred from the low
best-fit column density $n_{\mathrm{H}}$ that H1821+643 has a soft
excess.  However, the best-fit absorbed powerlaw is a sufficiently
good description of the readout streak spectrum so we have used this
simplistic model in our analysis.

The best-fitting model gave an unabsorbed flux of
$\left(1.45^{+0.04}_{-0.03}\right)\times10^{-11}\ergpcmsqps$ in the energy range
$2-10\keV$ and a reduced $\chi^{2}_{\nu}=1.6$.  The quasar flux
measured from the readout streak spectrum falls between the values
determined for this object by \citet{Fang02} and \citet{Yamashita97}.
This gives a $2-10\keV$ quasar luminosity of
$\left(4.2\pm0.1\right)\times10^{45}\ergps$.

This spectral model from the quasar readout streak was used as the
principal component in the ChaRT simulation of the quasar.  


\subsection*{A2 ChaRT simulations}
The \emph{Chandra} Ray Tracer (ChaRT), in conjunction with the
\textsc{marx} software, was used to simulate the \emph{Chandra} PSF
produced by the quasar.  Following the ChaRT analysis
threads\footnote{See http://cxc.harvard.edu/chart/}, the readout streak
spectrum and observation exposure time were used to generate a
ray-tracing simulation.  The output from ChaRT was then supplied to
\textsc{marx} version 4.4.0 which projected the ray traces onto the detector and
applied the detector response.  

ChaRT provides a user-friendly interface to the SAOTrace semi-empirical
model (\citealt{Jerius95}) which is based on the measured
characteristics of the mirrors, support structures and baffles.  The
model is then calibrated by comparison with actual observations.
However, while the PSF core and inner wing region match well with
observations (\citealt{Jerius02}), the PSF wings beyond $10\arcsec$,
produced by mirror scattering, seem to be underpredicted by the
raytrace model.  At the moment, the SAOTrace model does not model the
dither motion of the telescope or include residual blur from aspect
reconstruction errors.  \textsc{marx} includes the effects of these
errors using the DitherBlur parameter.  We have therefore tested
the use of SAOTrace to simulate the H1821+643 quasar spectra by
performing an identical analysis on the quasar 3C\,273, which has no
surrounding cluster emission.  Both of these
sources are close to on-axis and only the PSF core and inner wing
region are of interest to our analysis.  Beyond $10\arcsec$ the
cluster emission in H1821+643 will dominate and an accurate model of
the quasar PSF becomes much less important.

\subsubsection*{A2.1 Test object 3C\,273}
3C\,273 is a bright, nearby ($z=0.158$) radio-loud quasar with a jet
showing superluminal motion and has been intensively studied at
different wavelengths (eg. \citealt{Courvoisier98}; \citealt{Soldi08}).
3C\,273 was observed with \emph{Chandra} ACIS-S for $160\ks$ to study
the bright X-ray jet in detail (\citealt{Jester06}).  Here we used one
of the $40\ks$ \emph{Chandra} ACIS-S exposures of this object (obs. id
4879) to test whether a ChaRT simulation using the readout streak
spectrum would recover the correct result.  This observation of
3C\,273 was similar to that of H1821+643, featuring a piled up point
source, strong readout streak, bright PSF wings but crucially no
detected extended emission (the region containing the jet was
excluded).  Therefore this observation of 3C\,273 allowed us to test
our method of determining the quasar spectrum in a series of annuli
and compare the result with the observed spectra without the added
complication of superimposed cluster emission.

\begin{figure}
\centering
\includegraphics[width=\columnwidth]{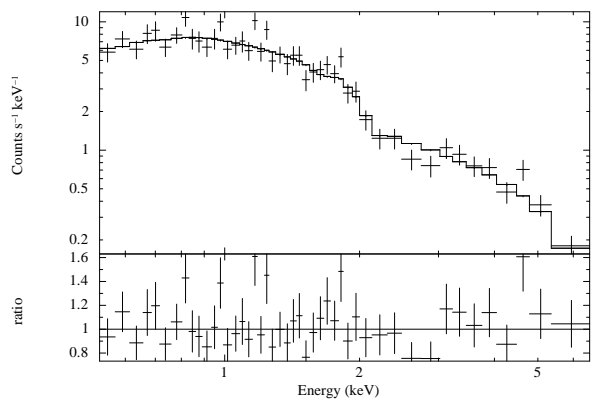}
\caption{3C\,273 readout streak spectrum and spectral fit.  The spectrum has
  been scaled by the ratio of the observation exposure time to the
  effective streak exposure time.}
\label{fig:3C273readout}
\end{figure}

As in section A1 for H1821+643, the readout streak spectrum was
extracted using two regions of the transfer streak either side of the
nucleus.  The spectrum was fitted with an absorbed power-law model
which provided a reasonable fit for our purposes over the required
energy range, reduced $\chi^{2}_{\nu}=1.2$.  This simple model gave a
best-fitting photon index $\Gamma=1.64^{+0.07}_{-0.04}$, Galactic
absorption upper limit of $n_{\mathrm{H}}<0.015\times10^{22}\pcmsq$ and unabsorbed flux
$\left(6.5^{+0.2}_{-0.4}\right)\times10^{-11}\ergpcmsqps$ in the
$2-10\keV$ energy range (Fig. \ref{fig:3C273readout}).  These
parameters fall within the expected range defined by the long term
variability of this source (\citealt{Chernyakova07}; \citealt{Soldi08}
and references therein).  Although 3C\,273 has also been found to have
a soft excess extending up to $2\keV$ (\citealt{Turner85};
\citealt{Staubert92}) we have ignored this for our simplistic
analysis.  3C\,273 has a slightly lower photon index than H1821+643
which produces a wider PSF because the PSF is energy-dependent (Figure
\ref{fig:SBprofile}).

\begin{figure}
\centering
\includegraphics[width=\columnwidth]{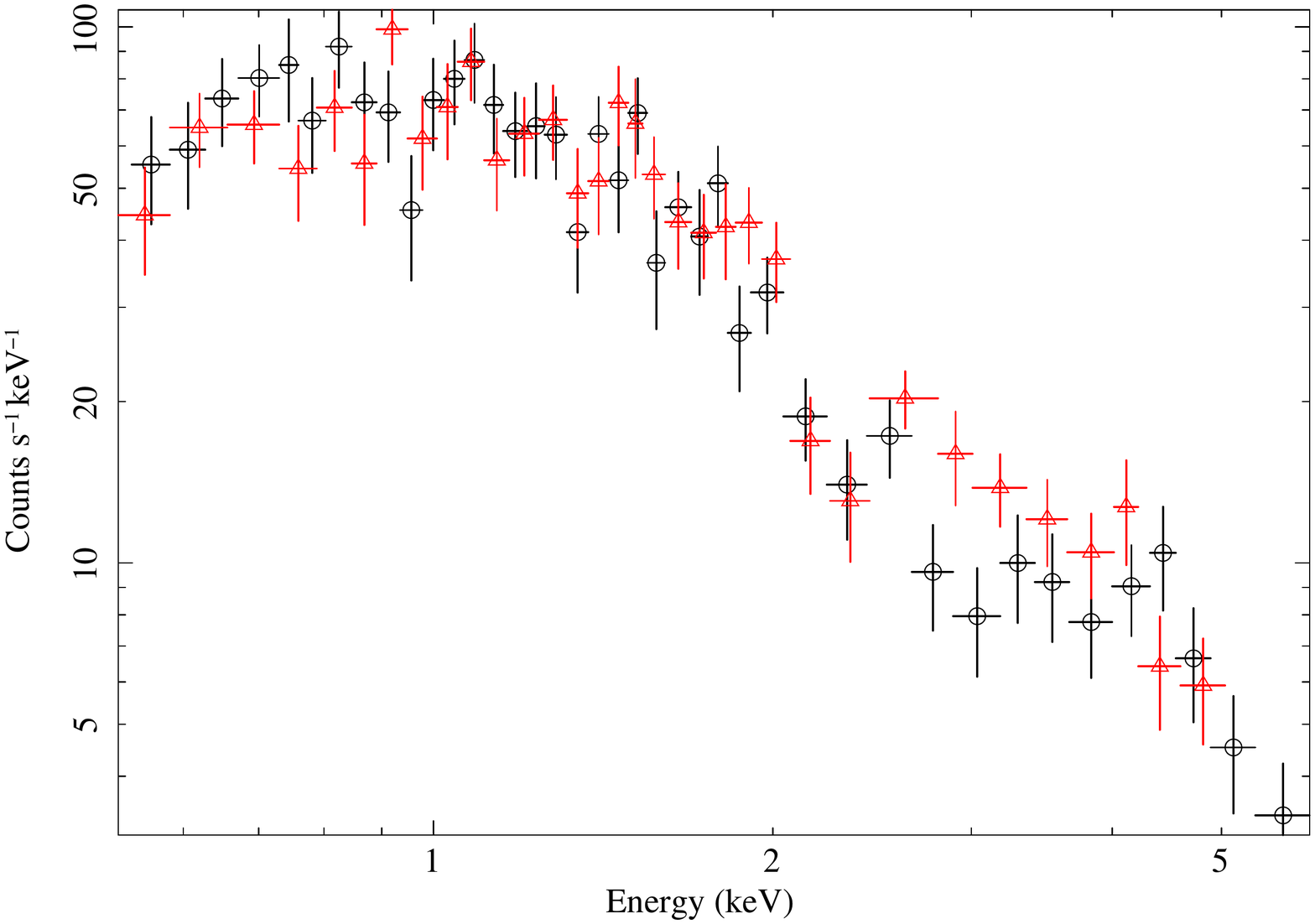}
\includegraphics[width=\columnwidth]{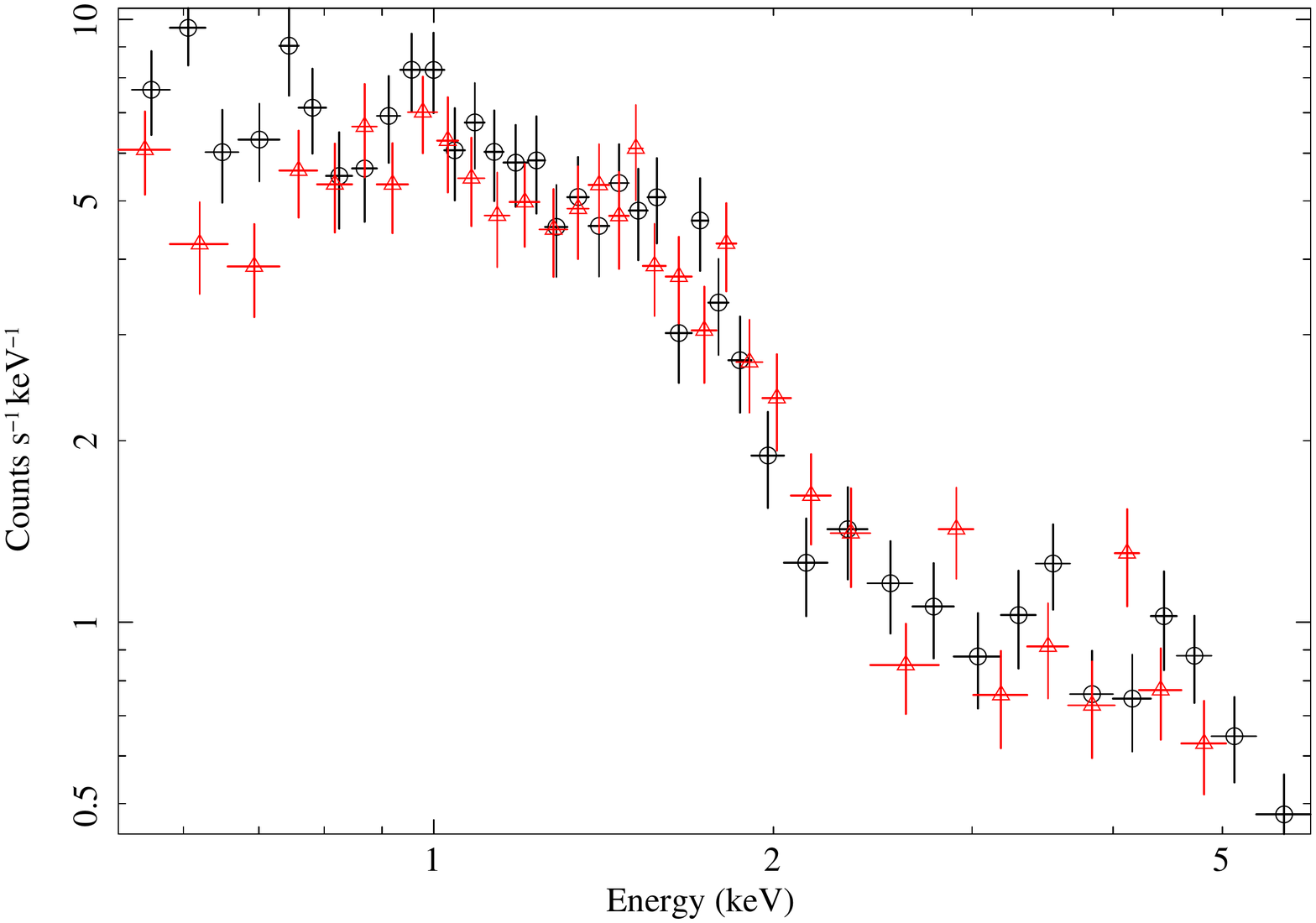}
\caption{Comparison of the observed deprojected spectra (black circles) and ChaRT
  simulated deprojected spectra (red triangles) for two annuli in 3C\,273.  Upper: $2.5-4.9\arcsec$
  Lower: $4.9-9.8\arcsec$.}
\label{fig:3C273compare}
\end{figure}

The best-fitting model to the readout streak spectrum was used as the
main input for the ChaRT simulations of 3C\,273.  Spectra were
extracted in two annuli, excluding the readout streak, with an
innermost radius of $2.5\arcsec$ outside of which pileup was
determined to be negligible.  The exposure time of each frame in this
observation of 3C\,273 was $0.4\s$, compared with $3.1\s$ for
H1821+643, which greatly reduced the number of piled up photons.  These
annuli were selected to contain the required minimum of 3000 counts
for a good deprojection and to cover a similar area of the PSF wings
to the regions used in the analysis of H1821+643.  The observed and
simulated spectra were deprojected and are shown overlaid in Fig.
\ref{fig:3C273compare}.  The simulated spectra were generally
well-matched to the observed spectra.  The mismatches can be mostly
accounted for by comparison with the model of the readout streak
spectrum, particularly at $\sim3\keV$ where the model was a
significant overestimate.  The best-fitting parameters for an absorbed
power-law model fitted to the observed and simulated deprojected
spectra were also found to be consistent within the errors.

A ChaRT simulation based on the readout streak spectrum was therefore
expected to provide an adequate description of the quasar spectrum for the series
of radial bins used in this analysis.  

\subsubsection*{A2.2 ChaRT Simulation of H1821+643}
A ChaRT simulation of the quasar H1821+643 was generated
for the same position on the detector using the model of the readout
streak spectrum.  ChaRT was run for an effective exposure time three
times longer than the total observation time to sample the range of
possible optical paths in the HRMA and reduce statistical errors.  The
raytraces were then projected onto the detector with \textsc{marx} to
create an event file.  Simulated quasar spectra were extracted
from the \textsc{marx} event file and combined with appropriate response
files for the
simulation\footnote{See http://space.mit.edu/CXC/MARX/examples/ciao.html}.

The width of the simulated PSF depends on the \textsc{marx} DitherBlur
parameter, which is a statistical term combining the aspect
reconstruction error, ACIS pixelization and pipeline pixel
randomization.  The default \textsc{marx} value is $0.35\arcsec$,
however the magnitude of the aspect blur is observation dependent.
The \textsc{marx} simulation of H1821+643 was repeated for DitherBlur
values of $0.25\arcsec$, $0.30\arcsec$ and $0.40\arcsec$ but for all
three cases the quasar-subtracted cluster surface brightness profile was
consistent within statistical errors.  In addition, all spectral
parameters were found to be consistent within errors to the
$0.35\arcsec$ simulation.  We therefore proceeded with the
DitherBlur parameter set to its default.

The simulated quasar spectra were extracted in identical regions to regions
1--11 (projected) and regions a--b (deprojected) shown in Fig.
\ref{fig:obsregions}.  The simulated quasar spectra from regions a--b were deprojected
with \textsc{dsdeproj}, as detailed in section \ref{deprojection}, to
ensure they have been processed in an identical way to the observed
deprojected spectra.  By determining the quasar spectral parameters from
the simulation for the same regions, we can fix these values in the
combined fit to the observation of the quasar and cluster.

The simulated quasar spectrum from each region was fitted with an
absorbed power-law model \textsc{phabs(powerlaw)} in \textsc{xspec};
the best-fitting parameters are given in table \ref{powvalues}.

The difference between the best-fit n$_H$ and power-law parameters
(Table \ref{powvalues}) and their input values from
the readout streak spectrum is caused by an incorrect calculation of the
effective area.  This also causes the variation in these parameters
between the radial bins.  The effective area correction for each spectrum,
provided by the ancillary response files (ARFs) assumes that all of
the PSF falls within that extraction region.  This was true for the
cluster spectrum in each annulus but not for the quasar spectrum,
where only the PSF wings were included (less than 5\% of the total
PSF).  The effective area correction for the quasar spectrum should
therefore be multiplied by the encircled energy fraction, which varies
as a function of energy.  When the ARF spectral response was
multiplied by the EEF calculated in a series of broad energy bands from the
ChaRT simulation, the input quasar parameters are recovered.



We confirmed our earlier analysis that pileup was not important beyond
a radius of $3\arcsec$ from the quasar by analysing a \textsc{marx}
simulation including pileup.  The pileup simulation produced
essentially identical results to the simulations without pileup and
all spectral parameters were consistent within errors.

\section*{Appendix B: Cloudy Simulations}
The most obvious way to model the intracluster gas using Cloudy is
with the \textsc{coronal equilibrium} command. However, this sets up a
constant temperature model which is not appropriate when we want to
consider the effect of the ionizing flux from the quasar on this
gas. 
Also, it is not clear that a pure cooling model is actually physically appropriate
since the cores of clusters are known to be deficient in cool gas and
this is widely thought to be due to the presence of a heat source
(eg. \citealt{PetersonFabian06}; \citealt{McNamaraNulsen07}). We
therefore chose to model the cluster emission by using an arbitrary
heat source and adjusting the heating rate to obtain the observed
temperature of the cluster through a balance of heating and
cooling. This was achieved using the \textsc{hextra} command.

Therefore, we first simulated cluster emission by creating a single
zone model heated using only the \textsc{hextra} pure heating command.
Metal abundances were set to match those measured by fitting a single
temperature mekal model to the X-ray spectra, $0.4\Zsun$. The heating
rate was adjusted until the temperature of the gas matched that
measured in the cluster in these regions ($15-30\kpc$ and
  $22-39\kpc$), $2.78\times10^7\K$ ($2.4\keV$). This model spectrum
was then renormalized to fit the cluster spectrum and the chi-square
statistic evaluated.  This Cloudy model for the cluster emission
provided a simple comparison for the Cloudy model with the added
quasar contribution; any differences could then be attributed to the
quasar input and not a discrepancy between the Cloudy and
\textsc{mekal} models of cluster emission.

To add in the quasar contribution we first generated the intrinsic
continuum shape for the active nucleus using the quasar SED model described in
section \ref{sec:accretion}.  To approximate the correct spectrum to
irradiate the cluster gas at the distance of our extracted X-ray
spectra we set up a Cloudy model with this continuum and
allowed it to pass though cluster gas at a temperature of
$2.78\times10^7\K$ ($2.4\keV$) for an equivalent hydrogen column density of
projected $2.3\times10^{21}\pcmsq$ ($15\kpc\times0.05\pcmcu$) or
deprojected $3.4\times10^{21}\pcmsq$ ($22\kpc\times0.05\pcmcu$).

The transmitted spectrum from the output of this model was then added
to the pure cluster simulation, allowing the \textsc{hextra} component
to vary so that the observed temperature of the cluster was still
reproduced, $2.78\times10^7\K$ ($2.4\keV$).  The emitted spectrum from
this Cloudy model together with a model accounting for the quasar PSF
(section \ref{sec:separation}) were fitted to the observed spectra
from the projected ($15-30\kpc$) and deprojected ($22-39\kpc$)
regions.

\bibliographystyle{mnras}
\bibliography{refs.bib}

\clearpage

\end{document}

%% file: defn.tex
\newcommand{\Mpc}{\rm\thinspace Mpc}
\newcommand{\kpc}{\rm\thinspace kpc}
\newcommand{\pc}{\rm\thinspace pc}
\newcommand{\km}{\rm\thinspace km}

\newcommand{\cm}{\rm\thinspace cm}
\newcommand{\pix}{\rm\thinspace pixel}
\newcommand{\yr}{\rm\thinspace yr}
\newcommand{\Gyr}{\rm\thinspace Gyr}
\newcommand{\Myr}{\rm\thinspace Myr}
\newcommand{\s}{\rm\thinspace s}
\newcommand{\ks}{\rm\thinspace ks}
\newcommand{\mus}{\hbox{$\mu\s\,$}}

\newcommand{\GHz}{\rm\thinspace GHz}
\newcommand{\MHz}{\rm\thinspace MHz}
\newcommand{\Hz}{\rm\thinspace Hz}

\newcommand{\K}{\rm\thinspace K}

\newcommand{\Msun}{\hbox{$\rm\thinspace M_{\odot}$}}

\newcommand{\Msunpyr}{\hbox{$\Msun\yr^{-1}\,$}}

\newcommand{\keV}{\rm\thinspace keV}

\newcommand{\erg}{\rm\thinspace erg}

\newcommand{\W}{\rm\thinspace W}

\newcommand{\ergpcmsqps}{\hbox{$\erg\cm^{-2}\s^{-1}\,$}}

\newcommand{\ergps}{\hbox{$\erg\s^{-1}\,$}}

\newcommand{\ergpspkpc}{\hbox{$\erg\s^{-1}\kpc^{-1}\,$}}
\newcommand{\WpHzpsr}{\hbox{$\W\pHz\psr\,$}}

\newcommand{\kmps}{\hbox{$\km\s^{-1}\,$}}

\newcommand{\kmpspMpc}{\hbox{$\kmps\Mpc^{-1}\,$}}

\newcommand{\plawnorm}{\hbox{$\rm\thinspace photons\keV^{-1}\cm^{-2}\s^{-1}\,$}}
\newcommand{\expmap}{\hbox{$\rm\thinspace photons\cm^{-2}\s^{-1}\pix^{-1}\,$}}
\newcommand{\Zsun}{\hbox{$\thinspace \mathrm{Z}_{\odot}$}}

\newcommand{\chisq}{\hbox{$\chi^2$}}

\newcommand{\asec}{\rm\thinspace arcsec}

\newcommand{\sr}{\rm\thinspace sr}

\newcommand{\emm}{\hbox{$\cm^{-5}\,$}}
\newcommand{\empasecsq}{\hbox{$\emm\asec^{-2}\,$}}
\newcommand{\keVempasecsq}{\hbox{$\keV\emm\asec^{-2}\,$}}

\newcommand{\pcmsq}{\hbox{$\cm^{-2}\,$}}
\newcommand{\pcmcu}{\hbox{$\cm^{-3}\,$}}

\newcommand{\pHz}{\hbox{$\Hz^{-1}\,$}}

\newcommand{\psr}{\hbox{$\sr^{-1}\,$}}

